# Obscured AGB stars in the Magellanic Clouds II. Near-infrared and mid-infrared counterparts[†]

Albert A. Zijlstra[1], Cecile Loup[2,3], L.B.F.M. Waters[2,4], P.A. Whitelock[5], Jacco Th. van Loon[1] and F. Guglielmo[6]

[1] European Southern Observatory, Karl-Schwarzschild-strasse 2, D-85748 Garching bei München, Germany
[2] Space Research Groningen, P.O.Box 800, NL-9700 AV Groningen, the Netherlands
[3] Institut d'Astrophysique de Paris, 98-bis Boulevard Arago, F-75014 Paris, France
[4] Astronomical Institute "Anton Pannekoek", University of Amsterdam, Kruislaan 403, NL-1098 SJ Amsterdam, the Netherlands
[5] South African Astronomical Observatory, P.O.Box 9, 7935 Observatory, South Africa
[6] Sterrewacht Leiden, P.O. Box 9513, NL-2300 RA Leiden, the Netherlands

**Summary.** We have carried out an infrared search for obscured AGB stars in the Magellanic Clouds. Fields were observed in the vicinity of IRAS sources with colours and flux densities consistent with such a classification. The survey uncovered a number of obscured AGB stars as well as some supergiants with infrared excess. We present photometry of the sources and discuss the colour diagrams and bolometric luminosities. One of the supergiants is close to the maximum luminosity allowed for red supergiants, implying a progenitor mass around $50\,{\rm M}_\odot$. Its late spectral type (M7.5) is surprising for such a massive star. Most of the AGB stars are luminous, often close to the classical limit of $M_{\rm bol} = -7.1$. To determine whether the stars are oxygen-rich or carbon-rich, we have acquired narrow-band mid-infrared photometry with the ESO TIMMI camera for several sources. All but one are found to show the silicate feature and therefore to have oxygen-rich dust: the colours of the remaining source are consistent with either an oxygen-rich or a carbon-rich nature. A method to distinguish carbon and oxygen stars based on $H-K$ versus $K-[12]$ colours is presented. We discuss several methods of calculating the mass-loss rate: for the AGB stars the mass-loss rates vary between approximately $5 \times 10^{-4}$ and $5 \times 10^{-6}\,{\rm M}_\odot\,{\rm yr}^{-1}$, depending on assumed dust-to-gas mass ratio. We present a new way to calculate mass-loss rates from the OH-maser emission. We find no evidence for a correlation of the mass-loss rates with luminosity in these obscured stars. Neither do the mass-loss rates for the LMC and SMC stars differ in any clear systematic way from each other. Expansion velocities appear to be slightly lower in the LMC than in the Galaxy. Period determinations are discussed for two sources: the periods are comparable to those of the longer-period galactic OH/IR stars. All of the luminous stars for which periods are available, have significantly higher luminosities than predicted from the period–luminosity relations.

## 1. Introduction

One of the least understood stages of stellar evolution occurs on the upper part of the Asymptotic Giant Branch (AGB). After the onset of thermal pulses, when the star alternates between helium- and hydrogen-shell burning, mass loss from the star starts to increase. Mass-loss rates as high as $10^{-4}\,{\rm M}_\odot\,{\rm yr}^{-1}$ have been found for Galactic AGB stars, causing the removal of almost the entire hydrogen envelope. Since the nuclear-burning rate on the AGB is less than $10^{-6}\,{\rm M}_\odot\,{\rm yr}^{-1}$, it is clear that mass loss dominates the evolution. After the high mass-loss phase ends, the star rapidly increase in temperature and evolves towards the planetary-nebula phase. All stars with initial masses less than $\sim 6$–$8\,{\rm M}_\odot$ (depending on metallicity) are expected to pass through this phase.

Knowledge of the AGB evolution is hampered by two major problems. First, no adequate theoretical formulation of mass loss exists, partly because the driving mechanism is not fully understood (for a review see Hearn 1989). Second, no accurate distances are known for the majority of evolved AGB stars. In the Milky Way, only the bulge provides a large sample of mass-losing AGB stars at fairly uniform distances. The stars in the outer bulge have been the subject of extensive studies (e.g. Whitelock et al. 1991), but this sample only contains old, low-mass stars and the distance spread is still quite large. The central part of the bulge may contain a sample of more massive stars at a known distance, but the large extinction makes these difficult to study. In contrast, the LMC provides a sample of AGB stars at a well-determined distance, which includes both low-mass and high-mass stars. It is for this reason that the study of AGB stars in the Magellanic Clouds is important.

Until recently, Magellanic Cloud studies were limited to optically visible AGB stars. This is a severe limitation, since it is well known that many Galactic AGB stars experience sufficiently high mass-loss rates to be optically obscured. The known LMC stars were therefore biased towards stars with low mass-loss rates, which have much less optical obscuration. Reid et al. (1990) compiled a catalog of IRAS sources aimed at finding mass-losing AGB stars. Using this catalog, Reid (1991) found evidence for the existence of obscured AGB stars in the LMC. He discovered a number of near-infrared sources very close to the IRAS position, in most cases without optical counterparts. In contrast, many of the suggested optical identifications of Reid et al. (1990) were at considerable distance from the IRAS source. Wood et al. (1992) also found a number near-infrared point sources without optical counterparts in the vicinity of IRAS sources, comprising both AGB stars and mass-losing supergiants. They showed from OH observations and period determinations that their sources were closely related to the Galactic OH/IR stars. The LMC is close to the detection limit of IRAS for AGB stars, and it is likely that in these studies only the most luminous objects were found. This is especially clear in the sample of Wood et al., who observed the brightest IRAS sources and found very luminous counterparts; whereas Reid (1991) finds $M_{\rm bol} = -5$ to $-6$, Wood et al. find for all their sources $M_{\rm bol} < -6$.

[†] Based on observations obtained at the European Southern Observatory and the South African Astronomical Observatory

We have used the catalogue of stellar IRAS sources in the Magellanic Clouds (Loup et al. 1995, paper I) to identify possible obscured AGB stars. The selection was done both on expected [12]–[25] IRAS colours and on flux density. Near-infrared imaging and photometry was obtained to identify which sources were obscured AGB stars. As was already found by Reid et al. (1990), the main confusion in the sample is caused by supergiants which however are easy to identify. In the following sections we present the source selection and observations, and discuss the photometry of the objects. In section 7 we discuss the results of of $10\,\mu$m imaging in narrow bands for five sources, from which we derive information on whether the stars are oxygen-rich or carbon-rich. Mass-loss rate determinations are discussed in section 9. Finally, for two objects with previous photometry, we discuss period determinations. Our results significantly enlarge the known sample of obscured AGB stars.

## 2. Source selection

There are two IRAS-based catalogues available for the Magellanic Clouds which go significantly deeper than the Point Source Catalog. The large catalogue of Schwering & Israel (1990) (see also Schwering & Israel 1989 and Schwering 1989) was derived from the pointed observations. Reid et al. (1990) used co-adding scans to increase the sensitivity and produced a catalogue for part of the LMC. Both used the Deep Sky Mapping data which is not used in e.g. the Faint Source Catalog, and as a result go deeper in the more crowded regions. (However, the FSC also covers the halo of the LMC and SMC which was not included in the Deep Sky Mapping program. Zijlstra et al. (1994) find for the planetary nebulae in the LMC detected by IRAS, that there is good agreement between the FSC and the Schwering & Israel catalogue, for those objects in common.)

Loup et al. (1995, paper I) have combined these catalogues to identify evolved stars with infrared excess in the LMC. They show that only AGB stars in the LMC with high mass-loss rates ($> 5 \times 10^{-6}\,{\rm M}_\odot\,{\rm yr}^{-1}$) and luminosity ($L > 10^4\,{\rm L}_\odot$) are within the detection limits of IRAS. Their selection of the most likely stellar sources in the IRAS database is based on IRAS colours, using the source-separation diagram of Pottasch et al. (1988), as well as on the expected flux levels. After applying their selection criteria and removing 131 foreground stars, they are left with more than 100 "best candidate" sources. From this list, we selected about 65 sources for infrared imaging and photometry. There is significant overlap with the sample of Reid (1991): of the 13 sources in his sample, 12 were independently also selected by us. Furthermore, there is one source in common with Wood et al. (1992). Some of the selected IRAS sources were identified with a bright star on the ESO-R Schmidt plates, making imaging observations unneccesary. In a number of cases we also found optical counterparts from a literature survey using SIMBAD. Because the source selection was done using a preliminary version of the catalog of paper I, there were still a few likely foreground stars in the sample.

At these low flux levels, the uncertainties on the IRAS fluxes can be the limiting factor on our ability to select AGB-star candidates. The uncertainties are discussed in more detail in paper I. The IRAS detection limit depends on the crowding: Schwering & Israel quote a general sensitivity limit of 0.15 and 0.22 Jy at 12 and 25 $\mu$m, respectively, but find a few sources with fluxes as low as 0.07 Jy. They note that the noise is mostly due to the LMC background and that the detectors themselves have lower noise. In addition, there are differences between the absolute calibrations of Schwering & Israel and Reid et al., with the latter being 25% lower at 25$\mu$m. From a comparison between sources in common between these two catalogues and the PSC, Loup et al. (paper I) derive a typical uncertainty on $C_{21} = \log(12 S_{25}/(25 S_{12}))$ of 0.25, corresponding to 1.4 mag, largely systematic and mostly due to the 25-$\mu$m flux. At 12$\mu$m, the statistical error varies from 1 mag for the faintest source to 0.2 mag for the remainder. The 60-$\mu$m and 100-$\mu$m fluxes are affected by confusion in some regions with high background flux, leading to flux ratios normally associated with HII regions: this may have eliminated some true AGB stars. The high success rate of the present survey indicates that the uncertainties are not disastrous, but complete samples cannot be obtained from the IRAS catalogues. It is not possible to derive uncertainties for individual sources.

Table 1 lists the selected sources, and where they were observed. Also listed is whether a counterpart was found on the ESO R-plates, labeled 'bright' if $m_{\rm R} \lesssim 10$. The star symbols indicate objects discovered by Reid (1991) and Wood et al. (1992). For each object we list the original name of the infrared source from Schwering & Israel (1990) or Reid (1990) in the first and second column. The third column gives the IRAS-PSC name, or if the object is not in the PSC we give the 1950 coordinates in the form normally used for IRAS PSC sources but without the prefix. In the cases where we could identify a likely optical counterpart in the literature, its name is listed if it was known before IRAS, and the spectral type if known. The naming convention is as follows: IRAS source names are PSC (IRAS Point Source Catalog), TRM (Reid et al. 1990), LI-LMC and LI-SMC (Schwering & Israel 1990). Names of optical sources are HV (Harvard Variable star, Payne-Gaposchkin & Gaposchkin 1966), WOH (Westerlund et al. 1981: "SG" for suspected supergiant, "G" for suspected giant), RMMP (Re-beiroth et al. 1983), SP (Sandulaek & Philip 1977), SK (the objective-prism survey of Sandulaek 1970). The names used in the remainder of the paper are indicated in boldface.

## 3. Observations

### 3.1. Near-infrared imaging

Near-infrared imaging of the selected sources was carried out on the MPI 2.2m telescope at La Silla on 2–6 January 1993. We used IRAC-2, which is a 256×256 NICMOS-3 infrared array, with standard ESO J,H,K filters. The weather was generally not photometric. During two nights cirrus appeared to be present, and the other two suffered from very high humidity. Some hours were lost due to humidity. The seeing was mostly $1''$ or better in the near infrared. The first three nights we used a pixel scale of $0.27''$/pixel, which gives a field of view of $69''$. However, this field may be insufficient for some sources as the uncertainty in the IRAS position can be large for faint sources in confused regions. On the fourth night we used a larger scale of $0.49''$/pixel, re-observing a number of fields where either the smaller field did not show an obvious counterpart or did not contain sufficient stars for astrometric purposes. A problem with the large scale is that the stellar images are undersampled, making it more difficult to extract reliable photometry.

Each field was observed for one minute per filter, followed by a one-minute exposure on-sky. For bright objects (including standard stars) the telescope was defocussed to prevent the array from becoming non-linear. Dome flats were taken for flat-fielding. In total about 50 fields were imaged. Most fields were found to contain an obvious, red object, in a few cases visible only at K. In other cases the infrared source was visible at J,H,K, but not on the finding chart which was made from the ESO Schmidt R-plates. Other objects were also bright on the R-plates. There were a few fields without a likely candidate. In some cases the red star is more than $30''$ away from the IRAS position and the association is not certain. Figure 1 shows as an example the IRAC-2 images at J and K of 05329−6709 (TRM60), one of the reddest objects in our sample.

### 3.2. Near-infrared photometry

Near-infrared (JHKL) photometry was carried out on 2–8 February 1993, using the Mk III photometer on the SAAO 1.9m telescope in Sutherland. A few additional observations were obtained during a run in March 1993. All sources detected with IRAC-2 were observed. In addition, sources with bright R-counterparts were observed, for which imaging was unnecessary, and a few IRAS sources for which counterparts had not yet been found. In the last case we attempted to find the counterpart by scanning the field using the K-filter. The SAAO photometer uses a dewar with an InSb detector and apertures of 3,6,9,12 or $18''$. If available, we used the IRAC-2 images to ensure that no other star was included in the aperture, or else to correct the derived magnitudes for the contaminating star. The photometry was calibrated using the standards HR0077 for the SMC and HR2015 for the LMC. All observations were transformed to the system defined by Carter (1990). The standard SAAO extinction law was assumed.

Observing conditions were generally good. One night was lost due to clouds, and during several other nights ridge cloud appeared towards the end of the night. Seeing was commonly $1-2''$ at the beginning of the night, significantly worsening at the end, especially before the appearance of ridge clouds. The aperture used was adjusted according to the seeing. All six usable nights were photometric.

### 3.3. Mid-infrared imaging and photometry

We used the ESO 10-$\mu$m camera TIMMI (Käufl et al. 1992) to observe six objects from our source list. The observations were carried out at the 3.6m telescope at La Silla on 1–3 December 1993. The conditions were photometric. TIMMI contains a 64×64-element gallium-doped silicon array with good cosmetic quality (one dead pixel) and a quantum efficiency of about 25%. We used the N-band filter ($\lambda_c = 10.10\mu m$, $\delta\lambda = 5.10\mu m$) as well as two narrow-band filters centered on the silicate and silicate-carbide features ($\lambda_c = 9.70\mu m$, $\delta\lambda = 0.49\mu m$; $\lambda_c = 11.30\mu m$, $\delta\lambda = 0.57\mu m$, respectively). The 40-mm lens yielding a scale of $0.45''$ per pixel and a field of view of $29''\times29''$ was used.

The normal observing procedure is to chop and nod so that two reference fields on both sides of the source field are continuously subtracted from the source field. The chopping frequency is several Hertz and nodding is done once or twice per minute. We selected a chop of $8''$, so that both reference frames would also contain the (point) source. This gives a significant improvement in signal-to-noise for point sources. In some cases one of the two reference frames could not be used because the source fell too close to the edge of the field. The telescope is essentially diffraction limited at $10\mu m$ (the first Airy ring has a diameter of $1.4''$) and we did not expect to resolve any of our sources. Integration times of approximately 45 minutes per source per filter were used. Flat-fielding was done by measuring a standard star at about ten positions on the array, and fitting a two-dimensional parabola to the measured values. The correction can reach a factor of two near the edges of the array but is less than 20% over most of the array. This was the reason why measurements close to the edge of the array were not used. The procedure does not correct for pixel-to-pixel variations and leaves a residual uncertainty which we estimate to be about 5%.

The final frames contain one 'positive' and two 'negative' sources, 17 pixels apart. The positive stellar image contains twice as much stellar light as each of the two negative stellar images. To derive reliable magnitudes, as well as reliable estimates of their accuracies, we derived 'magnitude profiles' by summing over a circular area and subtracting the locally-determined background, taking circles with a diameter increasing from 3 to 30 pixels. This was done for both program star and calibrator, and the two profiles were then subtracted. Ideally, this would yield a flat curve—i.e. the diameter of the circle should not affect the differential magnitude, as long as it is the same for program and calibration star. Partly because of seeing variations, in practice for each source we found a range of diameters within which the differential magnitude profile (hereafter DMP) was flat within the noise, with deviations occuring both close to the center of the stellar image as well as at large distances. The differential magnitude was derived by averaging the magnitude estimates in the flat part of the DMP. The advantage is that more accurate magnitudes are obtained and a reliable estimate of the accuracy is derived

**Table 1.** Selected sources and observing log

| LI-SMC/LMC | TRM | PSC (B1950) | IRAC-2 [a] | SAAO[a] | R-plate[b] | TIMMI[a] | optical | type | comment |
|---|---|---|---|---|---|---|---|---|---|
| LI-SMC 225 |  | PSC00165−7418 | y | y | b |  | **VV Tuc** | M4 | 1 |
| LI-SMC 005 |  | PSC00350−7436 | y | y | f |  |  | Ce+ |  |
| LI-SMC 185 |  | PSC01074−7140 | y | y | b | y | **HV12956** | M5e | 2 |
| LI-LMC1825 |  | PSC04286−6937 | y | y | n |  |  |  |  |
| LI-LMC1844 |  | PSC04374−6831 | y | y | n |  |  |  |  |
| LI-LMC0004 |  | PSC04407−7000 | y | y | n | y |  |  |  |
| LI-LMC0057 |  | PSC04496−6958 | y | y | f |  |  |  | 3 |
| LI-LMC0060 |  | PSC04498−6842 | y | y | f | y |  |  | 4 |
| LI-LMC0099 |  | PSC04518−6852 | n |  |  |  |  |  |  |
| LI-LMC0141 |  | PSC04539−6821 | y | y | n |  |  |  |  |
| LI-LMC0198 |  | PSC04557−6753 | y | y | f |  |  |  | 5 |
| LI-LMC0183 |  | PSC04553−6933 |  | y | b |  | **WOH SG071** | M2 |  |
| LI-LMC0203 |  | PSC04559−6931 | y | y | b |  | **HV12501** | M1.5 | 6 |
| LI-LMC0253 |  | PSC04581−7013 | y | y | b |  | **HV2255** | M4 |  |
| LI-LMC0273 |  | PSC04588−6811 | n |  |  |  |  |  |  |
| LI-LMC0297 |  | PSC05003−6712 |  | y | f |  |  |  |  |
| LI-LMC0310 |  | PSC05009−6616 |  | y | f |  |  |  |  |
| LI-LMC0463 | TRM009 | PSC05073−6752 | n |  |  |  |  |  |  |
|  | **TRM133** | 05079−6542 | n |  |  |  |  |  |  |
| LI-LMC0528 | TRM023 | PSC05099−6740 | y | y | f |  |  |  |  |
| LI-LMC0567 | TRM100 | PSC05110−6616 | n |  |  |  |  |  |  |
| ⋆LI-LMC0570 | TRM004 | PSC05112−6755 | y | y | n | y |  |  |  |
| ⋆LI-LMC0571 | TRM024 | PSC05112−6739 | y | y | n |  |  |  |  |
| ⋆**LI-LMC0578** | TRM072 | 05117−6654 |  | y | n |  |  |  |  |
| LI-LMC1880 |  | PSC05128−6455 | y | y | n |  |  |  |  |
| LI-LMC0671 |  | PSC05150−6942 | n |  |  |  |  |  |  |
| LI-LMC0732 |  | PSC05171−7048 | y | y | b |  | **HV928** | M3 | 7 |
| ⋆LI-LMC0793 | TRM020 | PSC05190−6748 |  | y | n |  |  |  |  |
| ⋆ | **TRM088** | 05203−6638 |  | y | n |  |  |  |  |
| LI-LMC1883 |  | PSC05208−6459 |  | y | f |  | **WOH G281** | M |  |
|  | TRM105 | 05215−6547 |  | y | b |  | **HD271191** | B0Iab | 8 |
| LI-LMC0932 | TRM108 | PSC05235−6544 |  | y | b |  | **HV12793** | M3/4 | 9 |
| LI-LMC0957 | TRM016 | PSC05242−6748 | n |  |  |  |  |  |  |
|  | TRM096 | 05270−6624 |  | n |  |  |  |  |  |
|  | TRM073 | 05276−6656 |  | y | b |  | **HV963** | M2/3I |  |
| LI-LMC1092 |  | PSC05278−6942 | n |  |  |  |  |  |  |
| LI-LMC1103 |  | PSC05281−6915 | y | y | b |  | **SP47-14** | M1 |  |
|  | TRM045 | 05283−6723 |  |  | n |  |  |  |  |
| **LI-LMC1116** | TRM114 | 05288−6731 |  |  | n |  |  |  |  |
| LI-LMC1130 | TRM099 | PSC05289−6617 | y | y | b |  |  |  |  |
| LI-LMC1137 |  | PSC05291−6700 | y | y | f |  |  |  |  |
| LI-LMC1157 |  | PSC05295−7121 | y | y | n | y |  |  |  |
|  | **TRM103** | 05298−6552 | n |  |  |  |  |  |  |
| LI-LMC1170 | TRM049 | PSC05299−6720 |  | y | b |  | **SP45-34** | M1Ia |  |
| LI-LMC1172 |  | PSC05300−6859 | y | y | b |  | **SP46-44** | M1Ia |  |
| ⋆LI-LMC1177 | TRM079 | PSC05300−6651 | y | y | f |  |  |  |  |
| LI-LMC1190 | TRM046 | PSC05304−6722 |  | y | b |  |  |  |  |
| LI-LMC1223 |  | PSC05313−6920 |  | y | b |  | **HV12998** | M1 | 10 |
| LI-LMC1234 | TRM089 | PSC05315−6631 | y | y | b |  | **HV990** | M2I |  |
| ⋆LI-LMC1238 | TRM101 | PSC05318−6604 | y | y | b |  | **WOH SG374** | M6 | 11 |
| LI-LMC1241 | TRM087 | 05318−6642 |  | y | b |  | **SP52-1** | M3/4I |  |
| ⋆**LI-LMC1259** | TRM112 | 05321−6744 |  | y | b |  |  |  |  |
| ⋆LI-LMC1286 | TRM060 | PSC05329−6709 | y | y | n | y |  |  |  |
| LI-LMC1304 | TRM063 | PSC05334−6706 | y | y | b |  | **HV5933** | M4I |  |
| LI-LMC1341 |  | PSC05346−6949 | n |  |  |  |  |  |  |
| LI-LMC1345 |  | PSC05348−7024 | y | y | n |  |  |  |  |
| LI-LMC1360 | TRM062 | 05354−6704 |  | y | b |  | **HV2700** | M1/2Iab |  |
| LI-LMC1366 | TRM068 | PSC05355−6657 |  | y | b |  |  |  |  |
| ⋆LI-LMC1382 | TRM077 | PSC05360−6648 | y | y | n |  |  |  |  |
| LI-LMC1602 | TRM135 | PSC05433−6728 | y | y | b |  |  |  |  |
| LI-LMC1765 |  | PSC05506−7053 | y | y | n |  |  |  |  |
| LI-LMC1768 |  | PSC05522−7120 | n |  |  |  |  |  |  |
| LI-LMC1780 |  | PSC05540−6533 | n |  |  |  |  |  |  |
| LI-LMC1790 |  | PSC05558−7000 | y | y | n |  |  |  |  |
| LI-LMC1795 |  | PSC05568−6753 | y | y | b |  |  |  |  |
| LI-LMC1803 |  | PSC05588−6944 | n |  |  |  |  |  |  |

[a]: y: detected, n: not detected, blank: not observed; [b]: b: bright, f: faint, n: not detected

**Table 1-continued.**

*comments to table 1*: 1. also known as HV6325. 2. Invisible on finding chart (R), but bright on TV screen. 3. AGB star is western component of double star. 4. Counterpart is eastern-half of very faint double, just visible on R-plate. 5. Just visible on R-plate. 6. Also known as RMMP61. Spectral type from Wood et al. (1983). 7. also known as ZZ Men, WOH SG215, RMMP269. 8. also known as SK−65 52. 9. also known as WOH SG257. 10. The HV star is also known as WOH SG369, RMMP519 and SP 47-22. There is a second star visible on R-plate much nearer the IRAS position, for which no photometry is available. 11. The star is south of the IRAS source, which was outside the IRAC field. It is possible that there is an obscured star nearer the IRAS position.

simultaneously: we use the standard deviation of the points with respect to the mean. We used MIDAS to process the images.

This technique was only applied to the 'positive' stellar images: the negative images were normally too close to the edge of the edge of the array to allow the calculations with larger circles. We therefore used the 'negative' stellar images as a cross-check only. Digital filtering is normally used in the standard TIMMI reduction software to remove high-frequency variations of the background: this gives an improvement in image quality due to the fact that the point-spread function is well sampled. However, it left the photometry as well as its accuracy essentially unaffected and we therefore did not apply the digital filtering.

As calibrator we used $\gamma$ Ret, for which we adopt magnitudes of $-0.71$ at N, $-0.73$ at $9.7\mu$m and $-0.77$ at $11.3\mu$m. (These values come from the (unpublished) ESO system; Bouchet et al. (1989) derive magnitudes which are about 0.05m fainter.) The N-band showed a significant atmospheric extinction term of 0.32 and 0.13 magnitude per airmass for the first and second night, respectively. For the $9.7\mu$m band, which coincides with strong atmospheric $O_3$ absorption, values of 1.4 and 0.6 magnitude per airmass were determined on the first and second night, respectively, and the $11.3\mu$m band showed an extinction of 0.26 magnitude per airmass for both nights. The estimated accuracy of the extinction is $\sim$20%: this is included in the uncertainties of the derived magnitudes. Our faintest source was detected at an N-band magnitude of about $7.2\pm0.1$ in 45 minutes, in good agreement with the expected sensitivity of $m_N = 6.3$ at ten sigma in one hour (Käufl et al. 1992).

## 4. Results

Table 2 lists positions for those objects for which a near-infrared counterpart was found close to the IRAS source, either on the IRAC-2 image or via scanning of the field. If an optical counterpart was identified, its position was measured on the ESO R-plate, or taken from the Guide Star Catalogue for stars included therein. For stars not detected on the R-plates, the position was determined from the IRAC-2 image relative to field stars. The sources found through field scanning have positions accurate to $\sim 10''$. For the other sources the 1-sigma accuracy is about $0.5''$. Table 2 lists the positional difference between the near-infrared and the IRAS source. In three cases the difference is so large that the association is doubtful.

For those sources in common with Reid (1991), the differences between our and Reid's positions are listed in Table 2. The positions are generally in good agreement. Differences around $10''$ are found for 04496−6958 and 05112−6739. There is poor agreement for one source only: LI-LMC1259 (TRM112), where the difference is $41''$ (mostly in right ascension). (In addition, LI-LMC0578 has a difference of $15''$, but for this source our position is not very accurate.) Reid suggests that the source may be extra-galactic in nature, but the difference in position makes it likely that an error was made in the identification of the optical counterpart. The object is located in a star-forming region and coincides with a bright knot on the ESO R-plate; it also shows no evidence for variability. We suggest it could be an HII region. One object (TRM045) we could not confirm, probably because of source variability.

The sources which were not detected are listed in Table 3. In some cases there are possible counterparts, which are not particularly red in the near infrared. There are several possible reasons for non-detections: (1) The objects lie outside the observed field. This is possible for a number of sources which were imaged with a $1'$ field. (2) The IRAS source is non-stellar (or non-existent). (3) The star associated with the IRAS source is neither intrinsically red nor significantly reddened. This could be the case for post-AGB stars having a detached envelope, although the IRAS flux will decrease rapidly after the envelope becomes detached. (4) The star is so highly reddened that the shell is optically thick at K. This is the case for at least one object: 05346−6946, which has been detected by Elias et al. (1986) at $10\mu$m. Further detections of such highly obscured objects will require imaging at longer wavelengths. The other cases could probably be resolved with ten-micron imaging. Detection of either case (3) or (4) would yield important additions to our sample.

In case of possible counterparts, we list the K-band magnitude as derived from IRAC-2. These are uncertain because of the non-photometric conditions. For sources without counterparts which were observed with the large field (so that it is unlikely that the counterpart was outside the observed field) we list an upper limit at K. This should be taken with caution: it only applies to a possible highly obscured star in the field. In case (3) above, the IRAS source would be associated with one of the detected stars in the field at considerably brighter magnitudes. For the sources observed with too small a field, no relevant upper limit can be given.

Table 4 lists the magnitudes, where in a few cases small corrections have been applied for confusing sources in the aperture. The IRAS flux densities are given in Janskys; they are not colour corrected. All near-infrared magnitudes are from the 1993 February SAAO observations, except for LI-LMC0578 which was observed in March and the magnitudes given in brackets which are derived from the IRAC-2 images. The SAAO JHK measurements are accurate to better than $\pm0.04$mag and the L measurements to better than $\pm0.06$mag, unless marked with a colon where the JHK are better than $\pm0.1$mag and the L better than $\pm0.12$mag. In the cases where the listed magnitudes are taken from the IRAC-2 data, the sources were too faint for photometric measurements. The IRAC-2 magnitudes are somewhat less reliable (also because no extinction correction could be applied due to the weather conditions), but at the faint levels the error is dominated by background photon noise rather than calibration uncertain-

**Table 2.** Measured positions

| name | NIR source RA DEC (J2000) | IRAS source RA DEC (J2000) | Δ (″) | Reid (1991) RA DEC (J2000) | Δ (″) | comments |
|---|---|---|---|---|---|---|
| VV Tuc | 00 18 53.7 −74 01 57 | 00 18 50.5 −74 02 15 | 23 | | | |
| 00350−7436 | 00 36 59.8 −74 19 50 | 00 37 00.3 −74 19 47 | 4 | | | |
| HV12956 | 01 09 2.32 −71 24 09 | 01 09 03.0 −71 24 08 | 4 | | | |
| 04286−6937 | 04 28 30.3 −69 30 49 | 04 28 30.6 −69 30 46 | 4 | | | |
| 04374−6831 | 04 37 22.8 −68 25 03 | 04 37 22.2 −68 25 17 | 14 | | | |
| 04407−7000 | 04 40 28.4 −69 55 12 | 04 40 28.0 −69 55 14 | 2 | | | |
| 04496−6958 | 04 49 18.6 −69 53 13 | 04 49 17.8 −69 53 13 | 4 | | | |
| 04498−6842 | 04 49 41.4 −68 37 50 | 04 49 40.8 −68 37 50 | 3 | | | |
| 04539−6821 | 04 53 46.3 −68 16 12 | 04 53 46.8 −68 16 26 | 14 | | | |
| 04557−6753 | 04 55 38.9 −67 49 10 | 04 55 38.1 −67 48 47 | 23 | | | |
| WOH SG071 | 04 55 03.2 −69 29 13 | 04 55 02.4 −69 29 13 | 4 | | | |
| HV2255 | 04 57 43.3 −70 08 50 | 04 57 43.7 −70 08 59 | 9 | | | |
| 05003−6712 | 05 00 19.1 −67 07 57 | 05 00 19 −67 08 03 | 6 | | | |
| 05099−6616 | 05 01 03.8 −66 12 40 | 05 01 04.1 −66 12 43 | 3 | | | |
| 05099−6740 | 05 10 02.8 −67 35 48 | 05 09 54.7 −67 36 42 | 47 | | | |
| 05112−6755 | 05 11 10.6 −67 52 11 | 05 11 10.2 −67 52 17 | 7 | 05 11 09.0 −67 52 07 | 10 | |
| 05112−6739 | 05 11 13.8 −67 36 30 | 05 11 13.4 −67 36 25 | 5 | 05 11 14.3 −67 36 22 | 9 | |
| LI-LMC0578 | 05 11 37 −66 51 00 | 05 11 41.2 −66 51 12 | 28 | 05 11 38.6 −66 51 12 | 15 | a,b |
| 05128−6455 | 05 13 04.6 −64 51 40 | 05 13 04.2 −64 51 39 | 19 | | | |
| HV928 | 05 16 38.4 −70 45 41 | 05 16 37.2 −70 45 35 | 8 | | | |
| 05190−6748 | 05 18 57 −67 45 00 | 05 18 56.2 −67 45 25 | 25 | 05 18 56.1 −67 44 55 | 7 | |
| TRM088 | 05 20 20 −66 35 45 | 05 20 21.0 −66 36 00 | 16 | 05 20 18.7 −66 35 47 | 8 | a,b |
| WOH G281 | 05 21 03.6 −64 56 43 | 05 21 03.3 −64 56 40 | 3 | | | a |
| HD271191 | 05 21 43.0 −65 44 57 | 05 21 42.2 −65 45 07 | 11 | | | b |
| HV12793 | 05 23 43.6 −65 42 00 | 05 23 43.1 −65 41 57 | 5 | | | |
| HV963 | 05 27 34.5 −66 53 31 | 05 27 36.5 −66 53 43 | 17 | | | |
| SP47-14 | 05 27 47.6 −69 13 21 | 05 27 46.6 −69 13 25 | 7 | | | b |
| 05289−6617 | 05 29 02.1 −66 15 27 | 05 29 02.0 −66 15 26 | 1 | | | |
| 05291−6700 | 05 29 07.6 −66 58 15 | 05 29 06.0 −66 57 48 | 28 | | | |
| 05295−7121 | 05 28 40.8 −71 19 13 | 05 28 49.2 −71 19 26 | 43 | | | |
| SP45-34 | 05 29 55.1 −67 18 37 | 05 29 55.0 −67 18 33 | 4 | | | |
| SP46-44 | 05 29 42.3 −68 57 18 | 05 29 42.8 −68 57 20 | 4 | | | |
| 05300−6651 | 05 30 04.0 −66 49 24 | 05 30 04.5 −66 49 04 | 20 | 05 30 03.5 −66 49 25 | 4 | |
| 05304−6722 | 05 30 21.0 −67 20 05 | 05 30 20.7 −67 20 14 | 9 | | | |
| HV12998 | 05 31 04.2 −69 19 04 | 05 31 01.5 −69 18 45 | 24 | | | |
| HV990 | 05 31 37.0 −66 30 08 | 05 31 36.5 −66 29 48 | 20 | | | |
| WOH SG374 | 05 31 47.5 −66 03 40 | 05 31 45.9 −66 03 51 | 15 | 05 31 47.4 −66 03 40 | 1 | |
| SP52-1 | 05 31 53.5 −66 40 43 | 05 31 55.4 −66 40 25 | 21 | | | |
| LI-LMC1259 | 05 32 03.4 −67 42 26 | 05 32 02.6 −67 42 28 | 5 | 05 31 56.3 −67 42 19 | 41 | |
| 05329−6709 | 05 32 51.9 −67 06 52 | 05 32 51.1 −67 06 56 | 6 | 05 32 51.1 −67 06 54 | 5 | |
| HV5933 | 05 33 26.9 −67 04 14 | 05 33 26.5 −67 04 21 | 7 | | | |
| 05348−7024 | 05 34 16.1 −70 22 53 | 05 34 15.6 −70 22 57 | 5 | | | |
| HV2700 | 05 35 19.0 −67 02 20 | 05 35 21.3 −67 02 43 | 27 | | | b |
| 05355−6657 | 05 35 28.3 −66 56 03 | 05 35 27.6 −66 56 06 | 5 | | | |
| 05360−6648 | 05 36 01.3 −66 46 40 | 05 35 59.5 −66 46 41 | 11 | 05 36 01.2 −66 46 38 | 2 | |
| 05433−6728 | 05 43 10.9 −67 27 28 | 05 43 11.6 −67 27 43 | 16 | | | b |
| 05506−7053 | 05 50 08.7 −70 53 12 | 05 49 57.7 −70 53 16 | 54 | | | |
| 05558−7000 | 05 55 21.1 −70 00 03 | 05 55 20.8 −70 00 05 | 2 | | | |

Comments
 a NIR position determined from SAAO K-scan (accuracy ∼ 10″)
 b IRAS position taken from Reid et al. (1990)

ties. One should keep in mind that there are uncertainties in the transformation between the Carter (SAAO) and ESO systems. The IRAS data is taken from Schwering & Israel (1990), or from Reid et al. (1991).

05329−6709, one of the reddest objects in the sample, appears strongly variable: the IRAC observations one month earlier indicated that it was more than one magnitude brighter at K. Although the conditions during the IRAC observations were not photometric, this cannot explain such a large difference as the standards only varied by 0.3m.

## 5. Source classification and sample definition

### 5.1. Present sample

Figure 2 shows the J−H,H−K diagram for the detected sources. This diagram gives a clear separation between (non-variable) M-giants and Miras (Feast et al. 1990). (The term 'M-giants' is used here for those RGB and AGB M-stars which are not undergoing strong Mira pulsations). The bottom panel shows an enlargement of part of the top panel; the boxes indi-

**Table 3.** Sources which are not detected or have uncertain counterparts

| Name | IRAS source RA | DEC (J2000) | $F_{12\mu m}$ (Jy) | $F_{25\mu m}$ (Jy) | possible counterpart RA | DEC (J2000) | K (mag) | comment |
|---|---|---|---|---|---|---|---|---|
| 04518−6852 | 04 51 39.8 | −68 47 29 | 0.37 | 0.22 | | | | 1 |
| HV12501 | 04 55 39.4 | −69 31 22 | 0.44 | 0.44 | 04 55 40.5 | −69 26 42 | 6.8 | 2 |
| 04588−6811 | 04 58 41.0 | −68 07 14 | 0.30 | 0.33 | 04 58 39.4 | −68 07 07 | 13.7 | 3 |
| 05073−6752 | 05 07 13.9 | −67 48 54 | 0.22 | 0.44 | 05 07 12.5 | −67 48 47 | 13.9 | 1,4 |
| TRM133 | 05 08 01.0 | −65 38 41 | 0.17 | 0.22 | 05 08 08.0 | −65 38 52 | | 1,5 |
| 05110−6616 | 05 11 10.8 | −66 13 03 | 0.19 | 0.33 | | | >14.0 | |
| 05150−6942 | 05 14 40.2 | −69 39 20 | 0.30 | 0.44 | 05 14 48.7 | −69 39 44 | 8.25 | |
| 05242−6748 | 05 24 08.2 | −67 45 42 | 0.19 | 0.22 | 05 24 08.3 | −67 45 48 | 13.5 | 6 |
| TRM096 | 05 26 54.6 | −66 21 04 | 0.12 | 0.08 | | | >12.5 | 7 |
| 05278−6942 | 05 27 23.2 | −69 39 44 | 0.37 | 0.44 | | | | 1,8 |
| TRM045 | 05 28 16.3 | −67 20 55 | 0.13 | — | 05 28 19.1 | −67 20 19 | >12.5 | 9 |
| LI-LMC1116 | 05 28 41.9 | −67 28 57 | 0.12 | 0.18 | | | >12.5 | 7 |
| TRM103 | 05 29 59.7 | −65 49 57 | 0.14 | 0.23 | | | | 1 |
| 05346−6949 | 05 34 14.2 | −69 47 21 | 7.40 | 21.09 | 05 34 13.7 | −69 47 29 | >15.5 | 10 |
| 05522−7120 | 05 51 32.0 | −71 19 34 | 0.19 | 0.22 | | | >15 | |
| 05540−6533 | 05 54 08.6 | −65 33 11 | 0.30 | 0.22 | 05 54 07.6 | −65 33 35 | 16.2 | |
| 05568−6753 | 05 56 38.8 | −67 53 40 | 0.33 | 0.44 | 05 56 42.6 | −67 53 20 | 10.8 | |
| 05588−6944 | 05 58 24.8 | −69 44 25 | 0.19 | 0.56 | 05 58 25.8 | −69 44 28 | 13.7 | |

Comments
1. Observed with too small IRAC-2 field (1′).
2. The IRAS position falls near a group of four stars, one of which is HV12501 (one of the others is classified as A3Ib); either of the four could be the counterpart. On the ESO R-plate there is a suggestion of nebulosity around the group, and the IRAS emission could well be associated with it rather than one of the individual stars.
3. There is a faint, red source at $\alpha = 04\ 58\ 47$, $\delta = -68\ 07\ 15$ which could also be the counterpart. However, the field is very crowded.
4. There is an additional faint, red source in the field which could be identified as the counterpart. However, the IRAS source is close to the edge of the 1′ field and the counterpart may have been missed altogether.
5. The IRAC field for TRM133 was centred at the stellar position given by Reid et al. (1990) which puts the IRAS position in fact outside the 1′ IRAC field. The near-infrared counterpart was detected by Reid (1991) and its position is listed in the table.
6. There is a star at the correct position, but the star is not obviously red.
7. Not observed with IRAC-2.
8. The field is extremely crowded and a faint counterpart could have been missed.
9. The counterpart was detected by Reid (1991). Scanning at SAAO did not recover the infrared source. Since we estimate the detection limit for scanning to be around K= 12.5, whereas Reid detected the counterpart at K= 11.17, we believe the source is variable. The listed position is from Reid.
10. The counterpart was detected by Elias et al. (1986) in the mid-infrared. However, they found it to be very faint in the near infrared (K= 16) and may have been below our detection limit. The luminosity indicates it is a supergiant.

cate the colour distribution of Galactic Miras and carbon stars (Feast et al. 1982, Bessell et al. 1989a, Whitelock et al. 1994), M-dwarfs (Tinney et al. 1993), M-giants (Feast et al. 1990) and LMC M-type supergiants (Wood et al. 1983). The colour systems used have been transformed to the Carter system using formulae given by McGregor (1994). Carbon stars tend to have a slightly redder J−H than (M-type) Miras but otherwise the two groups occupy the same general colour region. Both kind of stars can suffer from circumstellar reddening, which is the reason for the 'open-ended' box. The main confusion comes from supergiants: separating the M-giants and Miras from supergiants requires knowledge of the absolute magnitude. A few supergiants with very high circumstellar reddening are known in the Milky Way, so that even for reddened objects confusion between Miras and supergiants still exists.

Most of the stars in our sample separate into two different groups in the figure. One group has colours characteristic of M-giants, while the other group shows evidence for high reddening. Figure 3 shows the colour–magnitude diagrams as a function of the K-magnitude. It can be seen that the 'blue' group corresponds to the brighter stars at K, making an identification with supergiants or foreground M-giants likely.

A number of objects are rejected from the sample for various reasons. They are indicated as star symbols in Figures 2 and 3. The following three sources are rejected because of their colour and K-magnitude:

- LI-LMC1259. This object is located around (J−H,H−K)= (0.8,1.0) and falls outside any obvious group: this is the object we previously mentioned (Section 4) might be an HII region or young stellar object.
- 05433−6728 and 05289−6617. They appear close to the 'M-giant' clump while being much fainter at K. Based on the faint magnitudes, both objects are likely foreground stars and may not be related to the respective IRAS source.

There are in addition a number of objects which are likely Galactic (foreground) stars. They are:

- WOH G281. This object is a magnitude brighter at K than any of the others. From the VRI magnitudes, Westerlund et al. (1981) also suspect it is a foreground star. The 12$\mu$m flux is consistent with photospheric emission.
- VV Tuc. The association of the IRAS source with this known variable star, together with the very bright magnitudes which would give $M_{bol} = -8.5$ at the distance of the SMC, argues in favour of a foreground star although an identification as supergiant can not be completely excluded. There are no velocity measurements available.

**Table 4.** Magnitudes and IRAS flux densities

| Name | \multicolumn{4}{c}{Magnitudes} | | | \multicolumn{3}{c}{IRAS (Jy)} | |
|---|---|---|---|---|---|---|---|
| | J | H | K | L | 12 | 25 | 60 |
| VV Tuc | 8.34 | 7.36 | 7.12 | 6.79 | 0.52 | 0.33 | |
| 00350−7436 | 11.49 | 10.16 | 9.08 | 7.76 | 0.30 | 0.22 | |
| HV12956 | 10.96 | 10.14 | 9.72 | 8.88: | 0.41 | 0.44 | |
| 04286−6937 | (15.8) | 13.58 | 11.61 | 9.50: | 0.22 | 0.17 | |
| 04374−6831 | (16.7) | 14.27: | 12.01 | 9.28 | 0.19 | 0.17 | |
| 04407−7000 | 10.62 | 9.07 | 8.23 | 7.25 | 0.81 | 0.67 | |
| 04496−6958 | 12.10 | 10.19 | 8.80 | 7.25 | 0.37 | 0.33 | |
| 04498−6842 | 12.62: | 11.05 | 9.94 | 8.64: | 1.18 | 1.11 | |
| 04539−6821 | (18.2) | (14.7) | 12.19 | 9.14: | 0.19 | 0.22 | |
| WOH SG071 | 8.87 | 7.85 | 7.44 | 6.84 | 0.67 | 0.67 | |
| 04557−6753 | (15.9) | 13.33 | 11.34 | 8.75 | 0.26 | 0.22 | |
| HV2255 | 8.57 | 7.64 | 7.34 | 6.85 | 0.59 | 0.44 | |
| 05003−6712 | 12.36 | 10.96 | 9.78 | 8.33: | 0.44 | 0.44 | |
| 05009−6616 | 14.57: | 12.51 | 10.76 | 8.49: | 0.26 | 0.22 | |
| 05099−6740 | 14.62: | 12.85 | 11.54 | − | 0.19 | 0.44 | 0.8 |
| 05112−6755 | (17.2) | (15.0) | 12.68: | 9.43: | 0.41 | 0.33 | |
| 05112−6739 | − | (15.3) | 13.08: | − | 0.33 | 0.17 | |
| LI-LMC0578 | − | 12.54 | 10.67 | 8.35 | 0.15 | − | |
| 05128−6455 | 13.31 | 11.69 | 10.31 | 8.50: | 0.15 | 0.22 | |
| HV928 | 8.70 | 7.58 | 7.27 | 7.04 | 0.26 | 0.17 | |
| 05190−6748 | − | − | 12.16 | 8.63 | 0.34 | 0.23 | |
| TRM088 | − | 13.60: | 11.39 | 8.99 | 0.16 | − | |
| WOH G281 | 7.23 | 6.20 | 5.93 | 5.66 | 0.33 | 0.22 | |
| HD271191 | 7.77 | 7.01 | 6.82 | 6.46 | 0.18 | − | |
| HV12793 | 9.15 | 8.13 | 7.80 | 7.37 | 0.41 | 0.22 | |
| HV963 | 8.92 | 8.00 | 7.72 | 7.26 | 0.18 | 0.16 | |
| SP47-14 | 9.05 | 8.07 | 7.78 | 7.38 | 0.30 | 0.22 | |
| 05289−6617 | 12.43 | 11.43 | 11.22 | − | 0.11 | 0.44 | |
| 05291−6700 | 12.71: | 10.98 | 9.92 | 8.76 | 0.07 | 0.22 | |
| 05295−7121 | (15.2) | 12.89 | 11.03 | 8.94: | 0.19 | 0.17 | |
| SP45-34 | 9.08 | 8.16 | 7.88 | 7.50 | 0.19 | 0.22 | |
| SP46-44 | 8.02 | 7.20 | 6.99 | 6.64 | 0.26 | 0.11 | |
| 05300−6651 | (16.4) | 13.85: | 11.58 | 9.01: | 0.11 | 0.11 | |
| 05304−6722 | 8.78 | 7.99 | 7.69 | 7.06 | 0.48 | 0.22 | |
| HV12998 | 9.17 | 8.09 | 7.71 | 7.06 | 0.37 | 0.22 | |
| HV990 | 8.52 | 7.69 | 7.47 | 7.13 | 0.33 | 0.22 | |
| WOH SG374 | 9.97 | 9.24 | 8.80 | 7.82 | 0.36 | 0.35 | |
| SP52-1 | 8.83 | 7.90 | 7.65 | 7.30 | 0.11 | 0.17 | |
| LI-LMC1259 | 11.86 | 11.02 | 10.05 | 8.55 | 1.48 | 5.99 | |
| 05329−6709 | (16.6) | (13.2) | 11.65 | 9.20 | 0.85 | 1.83 | |
| HV5933 | 9.15 | 8.26 | 7.95 | 7.44 | 0.37 | 0.22 | |
| 05348−7024 | − | (19) | (13.8) | − | 0.48 | 0.28 | |
| HV2700 | 9.34 | 8.44 | 8.23 | 8.01: | 0.19 | 0.11 | |
| 05355−6657 | 8.35 | 7.47 | 7.24 | 6.81 | 0.33 | 0.22 | |
| 05360−6648 | (19) | (16.5) | 13.56: | − | 0.22 | 0.22 | |
| 05433−6728 | 12.08 | 11.02 | 10.90 | − | 0.15 | 0.22 | |
| 05506−7053 | (20) | (15.0) | 12.24: | 8.93: | 0.63 | 0.67 | |
| 05558−7000 | 11.80 | 9.93 | 8.84 | 7.46 | 0.74 | 0.67 | |

– HV928. Wood & Bessell (1985) find a velocity of about 60 km s$^{-1}$. It is identical with ZZ Men and is certainly a foreground star.

In the bottom panel of figure 3 it can be seen that the 'rejected sources' have on average significantly redder IRAS colours than the other stars. The two suggested M-giants are faint at 12$\mu$m and may be confused. The two possible foreground Miras, VV Tuc and ZZ Men, are much brighter and their association with the respective IRAS source is likely. Velocity measurements would be desirable for VV Tuc to establish its relation to the SMC.

There are three sources for which the colours could be consistent with unreddened Miras or even M-dwarfs: 05304−6722, WOH SG374 and HV12956. The first object we classify as supergiant, based on the bolometric luminosity. The second is identified with the optical star WOH SG374: Westerlund et al. (1981) classify it as a probable supergiant. The spectral type is M6 which would be unusually late for a supergiant. We find that the bolometric magnitude would be low for an LMC supergiant, but high for a Mira. The lack of reddening is surprising given the strength of the mid-infrared emission. HV12956 is a well-known Harvard variable (e.g. Whitelock et al. 1989) for which the association with the SMC has not been in doubt so far and which is known to be a long-period Mira variable. We later will show that the IRAS source is indeed associated with the star. Based on the present data we note the possibility that both sources might be foreground Miras. However, they will be included in the final sample.

**Table 5a.** MC IRAS-AGB stars from other sources

| name | J | H | K | L | L' | 12 | 25 | 60 | P | comment | spectral type |
|---|---|---|---|---|---|---|---|---|---|---|---|
| *Whitelock et al. 1989 (SMC)* | | | | | | | | | | | |
| 00483−7347 | 11.58 | 9.97 | 8.79 | 7.38 | − | 0.65 | 0.48 | | >1000 | | |
| 00554−7351 | (15) | 12.98 | 10.85 | 8.47 | − | 0.37 | 0.20 | | 800 | | |
| *Reid 1991 (LMC)* | | | | | | | | | | | |
| TRM045 | 16.18 | 13.36 | 11.17 | − | − | 0.13 | − | | | | |
| *Wood et al. 1992 (SMC)* | | | | | | | | | | | |
| 00477−7343 | 10.59 | 9.69 | 9.49 | − | 9.29 | <0.40 | 1.12 | 10.4 | | non-Mira? | |
| 00521−7054 | 13.86 | 12.53 | 11.23 | − | 9.07 | 0.28 | 0.80 | 1.02 | | | |
| 01039−7305 | 14.68 | 13.54 | 11.76 | − | 9.31 | 0.27 | 0.95 | 3.7 | | non-Mira? | |
| *Wood et al. 1992 (LMC)* | | | | | | | | | | | |
| 04509−6922 | 10.46 | 9.20 | 8.35 | − | 7.11 | 0.74 | 1.00 | | 1290 | | |
| 04516−6902 | 11.59 | 9.89 | 8.86 | − | 8.05 | 0.63 | 0.78 | | 1090 | | |
| 04530−6916 | 13.76 | 11.70 | 9.79 | − | 7.49 | 2.00 | 4.97 | | 1260 | | |
| 04545−7000 | 16.34 | 13.45 | 10.64 | − | 8.16 | 0.44 | 0.89 | | 1270 | | |
| 04553−6825 | 9.64 | 8.06 | 6.99 | − | 5.08 | 7.07 | 11.21 | | 930 | **WOH G064** | M7.5 |
| 05216−6753 | 12.84 | 12.02 | 10.38 | − | 8.16 | 3.91 | 12.39 | | | non-Mira? | |
| 05247−6941 | 8.62 | 7.49 | 6.96 | − | 6.46 | 1.04 | 2.77 | | | **WOH SG264** | M2 |
| 05261−6614 | 9.30 | 8.23 | 7.79 | − | 7.38 | 0.56 | 0.78 | | | | |
| 05294−7104 | 12.01 | 10.45 | 9.24 | − | 8.12 | 0.74 | 0.78 | | 1040 | | |
| 05298−6957 | 13.83 | 12.14 | 10.29 | − | 7.53 | 0.85 | 1.33 | | 1280 | | |
| 05389−6922 | 8.17 | 7.21 | 6.70 | − | 5.99 | 2.11 | 2.22 | | | **WOH SG453** | M0.5 |
| 05402−6956 | 14.58 | 12.02 | 10.12 | − | 8.14 | 0.63 | 0.89 | | 1390 | | |

**Table 5b.** Known LPVs in the LMC detected by IRAS

| name | | J | H | K | 12 | 25 | P | optical name | spectral type |
|---|---|---|---|---|---|---|---|---|---|
| 0453582−690242 | LI-LMC 143 | 10.25 | 9.29 | 8.67 | 0.11 | 0.17 | 857 | **WOH SG061** | M0 |
| 0454257−684856 | LI-LMC 153 | 11.39 | 10.47 | 9.68 | 0.30 | 0.17 | 728 | **SP 30-6** | M8 |
| 0510004−692755 | LI-LMC 530 | 12.63 | 11.81 | 11.51 | 0.22 | 0.22 | 169 | | ? |

For three objects the separation between the IRAS source and the near-infrared source is more than 30″, casting doubt on their association. They are: 05099−6740, 05295−7121, 05506−7053. In all three cases the near-infrared colours are indicative of significant reddening. Although IRAS sources sometimes show such large positional deviations, there is a possibility that these sources have been discovered by accident.

*5.2. Previously known stars*

In addition to our sample, candidates for IRAS-detected AGB stars in the Magellanic Clouds have been reported by Whitelock et al. (1989), Reid (1991) and Wood et al. (1992). Fourteen objects are listed in Wood et al. (1992) (excluding one which is also in our sample); their sample includes both AGB stars and supergiants. Three of their objects are in the SMC. There is one additional source in Reid (1991) (excluding the objects already in our sample and one object in common with Wood et al.). Two SMC IRAS-Miras are listed by Whitelock et al. (1989) (excluding one already in our sample). Table 5a lists the additional objects. The photometry is from the original sources and is in various photometric systems. Especially at J a significant correction between systems is required for red objects ( McGregor 1994); the conversion between the L and L' systems is also uncertain.

Three sources listed in Wood et al. are not included in Table 5a:

− 04571−6954 is identical with the B8Ia star HD 268835, also known as S73 or R66, and is a known S-Dor variable (Shore & Sandulaek 1984). Due to its early spectral type, the luminosity is far higherthan derived by Wood et al. based on the near-infrared flux.
− 05244−6832 is a very red IRAS source with little variability. For this reason Wood et al. suggest it might be a post-AGB star. However, it is very close to the HII region LHA 120-N 1380 and it is probably not an evolved star.
− 05325−6743 is identical to the well-known HII region LHA 120-N 57A. It also has the IRAS colours expected for an HII region.

Furthermore, Hughes & Wood (1990) have presented a catalog of optically visible long-period variables (LPVs) in the LMC: we have correlated our IRAS source list with this catalog which gave three further possible detections. They are listed in Table 5b. The first column lists the name as used in Hughes & Wood. The last two columns list optical identifications as well as spectral type. The first two objects have also been classified as (super)giants, but the periods and magnitudes are consistent with LMC Miras.

The additional sources in Tables 5a and 5b are indicated in Figure 3 by the cross symbols, where we have corrected the magnitudes to the Carter system. Figure 4 shows the J−H versus H−K diagram for these sources. The filled circles indicate the sources for which Wood et al. (1992) detect OH emission at velocities corresponding to the LMC. (One of these is not in Table 5a because it is also in our sample). There are a few sources for which the classification is not clear; in the tables these are indicated by a question mark:

- 01039−7305 and 05216−6735 are brighter at J than expected from the H−K colours, which may indicate a blue underlying star. The two objects are very similar: both have very red IRAS colours and show only small-amplitude variations. However, 05216−6735 is very bright and is classified as a supergiant, whereas 01039−7305 is much fainter and could be a post-AGB star.
- 00477−7343 could be a foreground star, based on its JHK colours, but it is also one of the brightest IRAS sources in the sample: the IRAS colours are characteristic of an HII region. The IRAS source is near a supernova remnant (0047-73.5) which together with the IRAS colours makes an association with a star-forming region likely.
- 0510004−692755 has a very short period. It also does not show noticeable reddening and the IRAS detection is therefore surprising. Further observations to confirm its association with the IRAS source would be desirable.

Separating supergiants from AGB stars requires knowledge of the luminosity. We will therefore first discuss the method used to determine the bolometric magnitude, before returning to the problem of classification.

### 5.3. Bolometric magnitudes

We calculate the bolometric magnitudes from the available infrared photometry, by integrating under a spline curve fitted to the J,H,K,L, 12-$\mu$m and 25-$\mu$m fluxes as function of frequency. At the blue end, we extrapolate a line joining the K flux to a point lying midway between the J and the H data points. At the other end, we assume zero flux at zero frequency. The procedure is discussed in detail in Whitelock et al. (1994) and it is better than a blackbody fit for sources with a significant fraction of the energy arising longward of L. Whitelock et al. show that the difference with blackbody fits is about $0.05^{\rm m}$ when J−K is between 1.1 and 1.8. The procedure underestimates the luminosity for extreme IRAS sources where a substantial fraction of the energy comes out longward of $25\mu$m. The reddest sources may have $F_{60\mu m} = 3 \times F_{25\mu m}$, leading to an underestimation by 25−50%. But is unlikely that AGB stars with such colours have been discovered in the LMC, because the $12\mu$m would for AGB stars be below the detection limit. Very red supergiants may be affected.

The accuracy of the derived bolometric magnitudes is limited by several effects. First, the measurement errors, although negligible for supergiants, are significant for the fainter AGB stars. For very red sources the total luminosity is dominated by the $25\mu$m flux, which for our sample is the most uncertain data point. An error of a magnitude for such a source could lead to an uncertainty of almost a factor of two in the luminosity. Second, there are strong water-vapour absorption bands for oxygen-rich M-stars in the near infrared, which fall between the standard JHKL bands and are difficult to account for in the flux determination. The overall effect of these results in the bolometric magnitudes being overestimated by 0.1 to 0.2m in optical Miras (Robertson & Feast 1981). The effect will be less in stars with thick dust shells. Third, variability will affect the bolometric magnitudes. The sources from Wood et al. have good period coverage, but the others have only been observed once. The average bolometric magnitude may differ from the ones derived here by typically a few tenths of a magnitude although larger differences are possible for the sources with largest amplitudes. Variability will especially affect the AGB stars: supergiants tend to have lower amplitudes.

For the sources taken from the literature we have redetermined $m_{\rm bol}$ using the present method: the agreement with the previously determined values is in general good to within 0.1 magnitude. In two cases the IRAS colours were so red that the present method would miss a significant part of the bolometric flux. For these objects (05216−6753 and 00477−7343) a literature value will be quoted instead. 00477−7343 has a large discontinuity between the near-infrared and IRAS colours, so that any fit is poorly constrained.

For conversion to luminosity, we assume a distance modulus for the LMC of $(m-M)_0 = 18.47 \pm 0.15$ and for the SMC $(m-M)_0 = 18.78$ (Feast & Walker 1987), corresponding to 49 and 57 kpc respectively.

### 5.4. Supergiants and peculiar sources

We will now attempt to distinguish between supergiants and Miras, using several criteria. The clearest indicator is luminosity: although there is a possible area of overlap in luminosity, supergiants are in general much brighter than AGB stars and one could classify objects with luminosities above the classical AGB limit of $M_{\rm bol} = -7.1$ as supergiants. Blöcker & Schönberner (1991) have shown that in some cases AGB stars can attain luminosities higher than the classical limit and this criterion should therefore be not be applied too strictly. For sources with only one observation, a Mira which happened to be near maximum may also have $M_{\rm bol} < -7.1$. We have chosen to classify all objects with $M_{\rm bol} < -8.0$ as supergiants. Two further criteria are used. Objects with $M_{\rm bol} < -7.1$ and small amplitude variability typical of supergiants ($< 0.5^{\rm m}$ at K) are classified as such. This information is available for the sources in Wood et al. (1992). Finally, objects with $M_{\rm bol} < -7.1$ and colours within region IV of Figure 2 are classified as supergiants. We added 04530−6916 to the list because it is very close to the chosen $M_{\rm bol}$ limit and its luminosity was probably underestimated because of the very red IRAS colours.

We note that it is possible that some objects are still misclassified. The best way to classify these objects would be through abundance determination, in particular s-process elements or lithium (Smith & Lambert 1989, 1990) which are dredged up in luminous AGB stars but not in supergiants. Smith et al. (1994) found for a sample of 112 optically-visible red giants that lithium is mainly found in stars within a narrow luminosity range ($-7.2 \lesssim M_{\rm bol} \gtrsim -6.0$), in agreement with the classical limits. It is not known whether this conclusion can be extended to obscured, mass-losing stars: abundance studies would be desirable for the present sample.

Table 6 lists the bolometric magnitudes and the K−[12] colour for the 20 objects classified as supergiants. In addition, three objects are listed for which the classification is not clear, for reasons mentioned in subsection 5.2. Figure 5 shows the spectral energy distribution for the stars we classify as supergiants and for the objects with uncertain classification. For a black-body spectrum with an effective temperature of 2500 K we expect K−[12] = 0.76. A blackbody of this temperature is illustrated in one of the spectra of Fig. 5: HD271191, for which the 12 μm flux appears to be dominated by photospheric emission. In general a blackbody is not expected to give a good fit to these late-type stars: the near-infrared colours are modified by strong molecular absorption features, particularly $H_2O$ and CO, as well as $H^-$ (Bessell et al. 1989b). The exact contribution of dust to the 12 μm flux is therefore uncertain. However, most of the supergiants exhibit a small but clear excess in the IRAS bands, over an extrapolation from the near-infrared colours, indicating mass loss. The mass-loss rates appear not to be very high since the near-infrared colours are in most cases not noticeably reddened. From the K−[12] colours we estimate $\dot{M} < 10^{-8}\,M_\odot\,yr^{-1}$ (see Sect. 8.2, but allowing for the fact that $V_{exp}$ may be several times larger for supergiants than for AGB stars). Two of the supergiants exhibit significant reddening indicative of much larger mass-loss rates: 04530−6916 and 05216−6753. The last object clearly shows the excess flux at J mentioned in subsection 5.2.

Table 6. Bolometric magnitudes for supergiants and sources with uncertain classification

| Name | $m_{bol}$ | K−[12] |
|---|---|---|
| *supergiants* | | |
| 04530−6916 | 10.7 | 7.09 |
| WOH G064 | 9.2 | 5.61 |
| WOH SG071 | 10.5 | 3.42 |
| HV 2255 | 10.3 | 3.12 |
| HD 271191 | 9.3 | 1.33 |
| 05216−6753 | 10.07[a] | 5.07 |
| HV 12793 | 10.9 | 3.09 |
| WOH SG264 | 10.1 | 3.56 |
| 05261−6614 | 10.9 | 3.64 |
| HV 963 | 10.8 | 2.23 |
| SP 47-14 | 10.9 | 2.81 |
| SP 45-34 | 10.9 | 2.50 |
| SP 46-44 | 9.7 | 1.72 |
| 05304−6722 | 10.4 | 3.09 |
| HV 12998 | 10.9 | 2.91 |
| HV 990 | 10.2 | 2.57 |
| SP 52-1 | 10.7 | 1.72 |
| HV 5933 | 10.9 | 3.15 |
| HV 2700 | 11.1 | 2.68 |
| 05355−6657 | 10.1 | 2.34 |
| WOH SG453 | 9.7 | 3.94 |
| *uncertain* | | |
| 00477−7343 | 11.8[a] | 6.45 |
| 01039−7305 | 12.7 | 6.93 |
| 0510004−692755 | 13.2 | 6.29 |

[a] $m_{bol}$ taken from Wood et al. (1992).

WOH G064 shows a peculiar, flat energy distribution. This source is also the most luminous object in our sample with $M_{bol} = -9.3$. The optical counterpart, discovered by Westerlund et al. (1981), has a very late spectral type of M7.5 (Elias et al. 1986). The flat energy distribution may indicate binarity or asphericity. The luminosity of this object is close to the upper limit where red supergiants occur (e.g. Chiosi & Maeder 1986, their figure 3), implying a progenitor mass of close to 50 $M_\odot$. It is surprising that such a massive star has evolved to such a late type. This is one of the most interesting sources in the sample.

Two of the sources with uncertain classification show a very large excess at the IRAS bands without significant reddening in the near infrared. The association of star and IRAS source is therefore in doubt. For the first source, 00477−7343, the IRAS source is probably an HII region. 01039−7305 is classified as uncertain because of the excess emission at J.

## 6. AGB stars

There are 34 sources in the LMC and 5 in the SMC which are good candidates for AGB stars. The spectral energy distributions of these sources are shown in Fig. 6. Most objects are significantly reddened even in the near-infrared. A few cases have IRAS flux densities higher than expected from the near-infrared data (especially 04498−6842 and 05294−7104) but this may be due to variability. The sources are listed in Table 7: we give the bolometric magnitudes, calculated as described in the previous subsection, the K−[12] colour which is an indicator for the mass-loss rate (Whitelock et al. 1994), $\lambda_{10}$ which is the effective heating wavelength of the incident radiation, as defined by Jura (1987; see also Whitelock et al. 1994), and the bolometric luminosity.

The derived luminosities are in almost all cases above $10^4\,L_\odot$. It is clear that only the upper part of the luminosity function of MC AGB stars is detected. The IRAC-2 survey would have found fainter sources if the circumstellar extinction were not very high but lower-luminosity objects with moderate circumstellar shells were evidently not detected by IRAS. This might be anticipated since the (present) luminous sources are already close to the detection limit in the IRAS data base. Our results show that there are some high-luminosity AGB stars in the LMC which experience heavy mass loss, however on the present data we cannot decide whether lower-luminosity stars reach the same mass-loss rates and become similarly obscured. It is possible that the recently started near-infrared sky surveys, such as the DENIS project, will find such objects.

The classical upper limit for the luminosity of an AGB star, derived from the Chandresekhar limit for the mass of the core together with luminosity–core-mass relation, corresponds to $\log L_\star = 4.75$. Blöcker & Schönberner (1991) have recently shown that highly convective, massive AGB stars can evolve to luminosities well above those predicted by the luminosity–core-mass relation, i.e. $\log L_\star > 4.75$, during envelope burning. The luminosities in our sample do not extend above the classical limit. However, it is possible that overluminous objects would be misclassified as supergiants since it is difficult to separate the two groups. In fact, there are theoretical indications that the core mass of AGB stars does not grow beyond 0.9–1.2 $M_\odot$ (Han et al. 1994). Observationally this is difficult to confirm: the only evidence comes from the masses of white dwarfs in young clusters (Weideman 1987, Vassiliadis & Wood 1993) which also do not extend to the Chandresekhar mass. If this reduced upper mass limit is true, the actual 'classical' limit for the luminosity of AGB stars would be closer to $M_{bol} = -6.9$ in which case some objects in our sample would be overluminous. We conclude that it will be very difficult to prove or disprove the Blöcker & Schönberner scenario based on the present sample.

Table 7. Candidate AGB stars

| Name | $m_{bol}$ | K–[12] | $\lambda_{10}$ | $\log L_\star$ [$L_\odot$] |
| --- | --- | --- | --- | --- |
| 00350−7436 | 12.1 | 4.28 | 0.69 | 4.6 |
| 00483−7347 | 11.6 | 4.66 | 0.77 | 4.7 |
| 00521−7054 | 12.7 | 6.40 | 1.44 | 4.3 |
| 00554−7351 | 12.7 | 6.02 | 0.98 | 4.3 |
| HV 12956 | 12.1 | 5.04 | 0.75 | 4.6 |
| 04286−6937 | 13.3 | 6.34 | 1.10 | 4.0 |
| 04374−6831 | 13.4 | 6.58 | 1.07 | 3.9 |
| 04407−7000 | 11.3 | 4.34 | 0.71 | 4.8 |
| 04496−6958 | 11.8 | 4.12 | 0.65 | 4.5 |
| 04498−6842 | 11.6 | 6.52 | 1.24 | 4.7 |
| 04509−6922 | 11.3 | 4.49 | 0.78 | 4.8 |
| 04516−6902 | 11.8 | 4.81 | 0.93 | 4.6 |
| WOH SG061 | 12.0 | 2.74 | 0.38 | 4.5 |
| 04539−6821 | 13.3 | 6.81 | 1.11 | 4.0 |
| SP 30-6 | 12.6 | 4.63 | 0.69 | 4.2 |
| 04545−7000 | 12.3 | 6.28 | 1.27 | 4.4 |
| 04557−6753 | 13.0 | 6.25 | 1.01 | 4.1 |
| 05003−6712 | 12.3 | 5.31 | 0.95 | 4.4 |
| 05009−6616 | 12.8 | 5.67 | 0.93 | 4.1 |
| 05099−6740 | 13.2 | 6.27 | 1.41 | 4.0 |
| 05112−6755 | 12.8 | 8.08 | 1.25 | 4.2 |
| 05112−6739 | 13.2 | 8.11 | 1.19 | 4.0 |
| LI-LMC0578 | 13.1 | 4.98 | 0.70 | 4.0 |
| 05128−6455 | 13.0 | 4.76 | 0.83 | 4.1 |
| 05190−6748 | 12.8 | 7.30 | 1.05 | 4.1 |
| TRM088 | 13.4 | 5.77 | 0.86 | 3.9 |
| TRM045 | 13.6 | 5.33 | 0.85 | 3.8 |
| 05291−6700 | 13.1 | 3.55 | 0.72 | 4.0 |
| 05294−7104 | 11.8 | 5.34 | 1.04 | 4.5 |
| 05295−7121 | 13.2 | 5.60 | 0.95 | 4.0 |
| 05298−6957 | 11.7 | 6.61 | 1.27 | 4.6 |
| 05300−6651 | 13.6 | 5.55 | 0.88 | 3.8 |
| WOH SG374 | 11.3 | 4.26 | 0.49 | 4.8 |
| 05329−6709 | 11.7 | 8.01 | 1.52 | 4.6 |
| 05348−7024 | 12.7 | 9.21 | 1.17 | 4.2 |
| 05360−6648 | 13.4 | 8.34 | 1.29 | 3.9 |
| 05402−6956 | 12.0 | 6.10 | 1.21 | 4.5 |
| 05506−7053 | 12.2 | 8.16 | 1.30 | 4.4 |
| 05558−7000 | 11.6 | 4.91 | 0.85 | 4.7 |

## 7. Ten-micron photometry and envelope chemistry

### 7.1. TIMMI imaging

AGB stars can either have carbon-rich or oxygen-rich envelopes, depending on their initial C/O ratio and how much carbon has been dredged up following thermal pulses. The chances of an AGB star becoming a carbon star increase with diminishing envelope mass. The mass-loss process is crucial; if high mass-loss rates occur it will reduce the time available for dredge-up and increase the likelihood of the AGB star ending oxygen rich. The mass-loss efficiency probably depends on both the progenitor mass and metallicity.

For our obscured stars the composition can in principle be derived from the mid-infrared emission from the circumstellar envelope. In the case of a carbon-rich shell, we expect an emission feature at 11.3$\mu$m caused by SiC grain mantles. An oxygen-rich shell would lead to a silicate feature (either in emission or absorption) at 9.8$\mu$m. This last feature is quite broad and can be very strong; if present it will also dominate the emission at 11.3$\mu$m. However, the stars are too faint for mid-infrared aperture photometry and require imaging through a 10-$\mu$m camera.

We selected six sources from the objects tentatively classified as AGB stars for 10-$\mu$m photometry using the TIMMI camera (Käufl et al. 1992). To separate the features, we observed through three filters: standard N and two narrow-band filters centred at the features (see subsection 3.3). The results are shown in Table 8. Listed are: the observed magnitudes, the corresponding N-band (10$\mu$m) flux, the IRAS 12$\mu$m and 25$\mu$m flux, $A_{10}$ (indicating the strength of the silicate feature), the K magnitude and the H−K colour. The last two are taken from observations at SAAO carried out in the first week of 1993 November, only four weeks before the TIMMI observations, thus minimising the effects of source variability. The N-band magnitudes are converted to fluxes assuming N = 0 corresponds to 37.0 Jy. The measured magnitudes are, in most cases, significantly different from those derived from IRAS fluxes. This may in part be due to the intrinsic uncertainties in the IRAS fluxes (caused by the large beam size and the background subtraction) which can be large for sources close to the IRAS detection limit in crowded areas. However, the difference in filters (N-band versus IRAS 12µm) as well as source variability may also make a contribution. For the largest-amplitude AGB stars, the 12µm flux can vary by as much as a magnitude (Le Bertre 1993). The results suggest that the Schwering & Israel catalogue should be used with care for such sources. In the case of 05295−7121, the poor agreement of the magnitudes plus the fact that the position of the IRAS source is almost 45″ removed from the near-infrared star suggests that the two are not related, and that the discovery of the AGB star was fortuitous. This source could not be observed in the two narrow bands because of its faintness.

Table 8. TIMMI results

| source | N | ± | magnitudes 9.8µm | ± | 11.3µm | ± | flux N (Jy) | IRAS (Jy) $F_{12}$ | $F_{25}$ | K | H–K | $A_{10}$ |
|---|---|---|---|---|---|---|---|---|---|---|---|---|
| HV12956    | 5.78 | 0.04 | 4.77 | 0.07 | 4.98 | 0.04 | 0.18  | 0.35 | 0.42 | 9.95  | 0.44 | 1.9   |
| 04407−7000 | 4.76 | 0.02 | 3.89 | 0.04 | 4.03 | 0.04 | 0.46  | 0.81 | 0.67 | 8.84  | 1.53 | 1.6   |
| 04498−6842 | 3.67 | 0.02 | 3.17 | 0.05 | 3.14 | 0.03 | 1.26  | 1.18 | 1.11 | 9.94  | 0.65 | 0.88  |
| 05112−6755 | 5.56 | 0.05 | 5.32 | 0.19 | 5.56 | 0.06 | 0.22  | 0.41 | 0.33 | 13.21 | 1.9  | 0.36  |
| 05295−7121 | 7.17 | 0.10 |      |      |      |      | 0.050 | 0.19 | 0.17 | 10.49 | 1.82 |       |
| 05329−6709 | 3.91 | 0.02 | 3.93 | 0.08 | 3.52 | 0.03 | 1.01  | 0.85 | 1.83 | 9.65  | 2.34 | −0.16 |

Of the five sources with sufficient measurements, three are much fainter in N than the narrow bands. Two of these are brightest at the silicate band, and appear to show strong silicate emission. The third source is somewhat brighter at 11.3µm which makes it difficult to classify. Of the two remaining sources, 05112−6755 is tentatively classified as having weak silicate emission, but the uncertainty on the magnitude is too large to be sure. Finally, 05329−6709 is brightest at the 11.3µm band and could either show SiC emission or it could be partly self-absorbed silicate: when the feature begins to be absorbed between 9 and 10µm it can still be slightly in emission longward of 10µm.

The classification can be secured by comparing to spectra of Galactic AGB stars. We have extracted from the IRAS Low Resolution Spectra (LRS) database (maintained in STAR-CAT) a random sample of spectra from classes exhibited by AGB stars. We subsequently multiplied the spectra with the passbands of the three filters; the same procedure was applied to the LRS spectrum of $\gamma$ Ret to calibrate the results. The resulting colours [N]−[9.8] versus [9.8]−[11.3] are plotted in Figure 7 for the various classes. We remind the reader of the following definitions: class 2n: blue continuum with silicate emission; 3n: blue continuum with silicate absorption; 4n: blue continuum with SiC emission; 6n: red continuum with silicate emission; 7n: red continuum with silicate absorption. The index n indicates the strength of the feature. In a few cases the spectra were obviously misclassified and have been reclassified.

The figure shows that the stars with SiC emission occupy a well-defined region as defined by these colours. The silicate stars show a very large scatter, because the silicate feature is much stronger than the SiC band and can be either in emission or absorption. Although a small number of the silicate stars coincide with the SiC stars, the chosen filters allow one to identify oxygen stars with confidence; however, a carbon star would not have unique colors. The five stars observed with TIMMI in all three bands are plotted with error bars. Two conclusions can be drawn: four out of the five stars definitely fall in the oxygen-rich category, and two sources have an unusually large ratio of [9.8] to [11.3] and of [9.8] to N, which could be indicative of a very strong silicate emission feature. One of the two, HV12956, is a possible Galactic foreground object. There are a number of stars with similar colours in the LRS database; they are invariably classified as '69' and are faint, compared to the other LRS sources, with generally $f_{12} \leq 20$ Jy. The continuum underneath the silicate feature is not detected in these stars which is the reason for the '69' classification. It is possible that these MC stars show a similar large silicate-to-continuum ratio, albeit at a much higher absolute flux. Alternatively, it is possible that the silicate feature is narrower in the Magellanic Clouds than in the Galaxy. 05112−6755 also has an unusual position in the diagram, but the error bars are large. For 05329−6709 the diagram is inconclusive. Although it is positioned right in the center of the distribution of carbon rich stars (class 4n), oxygen-rich stars can be found in the same region too.

Wood et al. (1992) also prove that several of their AGB stars in the LMC are oxygen rich, based on the discovery of OH maser emission. Their sources are generally more luminous than the present sample. 05329−6709 is also in their sample, and they detect OH maser emission from this source. Self-absorbed silicate emission is a common feature of heavily obscured OH/IR stars and its mid-infrared colours are thus consistent with an OH detection. This support its classification as an oxygen-rich source. (However, it is possible that the OH detection refers to a different source, since there are other IRAS sources within the Parkes primary beam, e.g. TRM58). We conclude that at least four and probably all five stars observed with TIMMI are oxygen rich.

The strength of the 9.8µm silicate feature is quantified by Schutte & Tielens (1989) through the parameter $A_{10}$, which is the magnitude excess at 9.8µm ($F_{\rm obs}$) compared to a continuum ($F_{\rm c}$) fitted between two narrow-band filters at 12.5µm and 8.7µm:

$$A_{10} = 2.5 \log(F_{\rm obs}/F_{\rm c}). \qquad (1)$$

$A_{10}$ can be estimated for our sources from the [N]−[9.8] colour. To show this, Figure 8a relates the two quantities for a number of randomly selected LRS spectra with classification 2n or 6n (i.e. silicate in emission). A reasonable correlation is found up to $A_{10} = 1.2$:

$$A_{10} = 2 \times ([{\rm N}] - [9.8]) - 0.12 \quad \pm 0.15. \qquad (2)$$

The resulting values are listed in Table 8. Figure 5 in Schutte & Tielens shows the predicted and observed relations between $A_{10}$ and the [3.8]−[12.5] colour temperature for a number of Galactic sources. We can compare their predictions with our sources by converting K−[N] to a colour temperature, which for our sources is in the range 600–1000K. The comparison is shown in Figure 8b, where the triangles present the data for Galactic stars from Schutte & Tielens, and the filled circles are the LMC stars. The line illustrates one model with a dust formation temperature of 750 K, taken from Schutte & Tielens, their figure 5a. They present a series of models which cover the observational points. Four of our sources fall above the models; the remaining one (05329−6709) is in good agreement. The two bluest LMC stars appear to have larger $A_{10}$ than Galactic AGB stars of similar colour temperature, whereas the two reddest LMC stars lie near the upper envelope for the Galactic stars. Based on the present sample, it appears possible that the emission strength of the silicate feature tends to be larger in the LMC than in the Galaxy. More mid-infrared observations of LMC stars will be necessary to confirm this. A stronger silicate feature can be caused by a *lower* silicate optical depth, since the feature peaks already at $\tau \approx 0.2$ (Groenewegen, private communication). Self-absorption reduces the strength at larger optical depth. The indication that the silicate feature is stronger in the LMC could thus be consistent with expectations based on the LMC metallicity.

We find that the present sample shows silicate features indicative of intermediate-to-low mass-loss rates, with $\tau_{9.8} < 3$.

### 7.2. Separating oxygen-rich from carbon-rich stars

A large number of AGB stars has been found in our Galaxy from follow-up studies of IRAS sources. Fouqué et al. (1992) and Guglielmo et al. (1994) observed more than 1300 IRAS sources and found in this way over 600 AGB stars. The stars can be easily separated into carbon stars and oxygen stars using the LRS classification. We have used this sample for comparison with the MC stars.

Figure 9 shows an H−K versus K−[12] diagram for the Galactic AGB stars. A clear separation between carbon and oxygen-rich stars can be seen, with the carbon-rich stars showing significantly redder H−K at the same K−[12]. The distinction holds up to K−[12]= 6. The main reason for the difference between carbon-rich and oxygen-rich stars in this diagram is that in oxygen-rich Miras the silicate feature increases the 12μm flux. The distinction is no longer clear once the circumstellar extinction becomes so large that the silicate band is no longer in emission. The stars with 9.8μm in absorption fall below the carbon stars.

For the sources observed with TIMMI, we can compare the classifications obtained from this diagram with our measurements. This requires a small correction factor since the N-band is not identical to the 12μm band. We have calculated the correction factor for each of the LRS spectra in Figure 8: we derive N−[12] = 0.3 ± 0.1 with no evidence for a colour term. This is only a minor correction to the plot.

Four sources (HV12956, 04407−7000, 04498−6842 and 05112−6755) fall clearly on the oxygen-rich sequence. 05329−6709 falls on the extreme end of the relation where most Galactic stars are carbon stars but a few oxygen stars are also found; on the basis of this diagram the classification is inconclusive. Wood et al. (1992) detected OH emission from this source which suggests that the position of the object in the diagram is due to the silicate feature becoming self-absorbed. The classification derived from the diagram gives good agreement with those obtained from narrow-band photometry. We therefore suggest that the same relation as found for Galactic stars can be used for stars in the LMC.

The use of the relation is not so easy for the MC sources not observed with TIMMI. The intrinsic uncertainties of the IRAS magnitudes, and to a lesser extent the variability, lead to estimated errors of ±1 magnitude in K−[12]. However, the diagram suggests that the objects with lower H−K tend to be oxygen rich. Ground-based mid-infrared photometry will be needed to apply this diagram to the faint LMC sources with more confidence. The success for the present sample indicates that for most of the sources only N-band photometry will be required.

05295−7121 falls below the relation, in a region avoided by Galactic AGB stars. For this object narrow-band photometry is not available, but the colours show it to be underluminous at 12μm. The reason is not clear. It is possible that it has the silicate feature in absorption, or that it has considerable interstellar reddening rather than circumstellar. In either case, the object may be misclassified.

### 7.3. Lack of carbon stars?

Oxygen-rich AGB stars are known to extend to the highest bolometric magnitudes known for MC AGB stars ($M_{\rm bol} = -7.1$). In LMC globular clusters, carbon stars become predominant above a bolometric magnitude of −4.8 or $\log L_* = 3.8$. Blanco et al. (1980) showed that this is also seen in the LMC bar. Lloyd-Evans (1984) has shown that the transition from M→C occurs at higher luminosity in younger clusters. He also shows that the transition occurs abruptly, with very few weak-banded carbon stars. In the LMC the age and metallicity content of globular clusters are well correlated, and his results therefore indicate that younger, more metal-rich stars reach higher bolometric magnitudes as oxygen stars. The highly luminous, oxygen-rich AGB stars in our sample would then belong to such a young and relatively metal-rich population, with high-mass progenitors. Further observations will be required to see whether this applies to the full sample. It would be very interesting to find carbon stars among the obscured AGB stars in the LMC, and to investigate whether a lower luminosity cut-off for such stars exists.

It is noticeable that there is only one luminous ($\log L_* = 4.3$) mass-losing carbon star known in the Magellanic Clouds (00554−7351, Wood et al. 1992, Whitelock et al. 1989). It is possible that the near-infrared observations were not sensitive to mass-losing carbon stars, since carbon grains are more effective in absorbing stellar light than silicate (oxygen) grains. The detection limit at K of the survey was not much better than K = 13 mag, and there could be AGB stars in the Magellanic Clouds with significantly higher circumstellar extinction, about which we have no information as yet. Imaging observations at longer wavelengths are required to find such stars.

## 8. Mass-loss rates

Many different methods have been used to estimate mass-loss rates for evolved stars. They all suffer from the problem that the observed species (e.g. CO, dust) only constitutes a minor fraction of the envelope and therefore requires a large correction factor which is not always well known. A good review of the available methods can be found in van der Veen & Olofsson (1989).

**Table 9.** Dust mass-loss rates

| Name | mass-loss derived from: | | | | | |
|---|---|---|---|---|---|---|
| | (K−[12]) | (L−[12]) | $A_{10}$ | H−K | K−L | OH |
| 00350−7436 | −7.91 | −7.25 | | −7.97 | −8.17 | |
| 00483−7347 | −7.69 | −7.09 | | −7.77 | −8.01 | |
| 00521−7054 | −6.70 | −6.51 | | −7.80 | −7.59 | |
| 00554−7351 | −6.92 | −6.86 | | −7.28 | −7.48 | |
| HV12956 | −7.48 | −6.54 | −7.73 | | | |
| 04286−6937 | −6.74 | −6.52 | | −7.51 | −7.78 | |
| 04374−6831 | −6.82 | −6.74 | | −7.41 | −7.52 | |
| 04407−7000 | −7.86 | −7.00 | −7.95 | | −8.83 | |
| 04496−6958 | −8.00 | −7.48 | | −7.58 | −7.88 | |
| 04498−6842 | −6.64 | −5.95 | −6.31 | −7.85 | −8.11 | |
| 04509−6922 | −7.79 | −7.09 | | — | −8.12 | |
| 04516−6902 | −7.61 | −6.64 | | −8.07 | | |
| WOH SG061 | −8.79 | | | | | |
| 04539−6821 | −6.47 | −6.79 | | −7.27 | −7.40 | |
| 04545−7000 | −6.77 | −6.77 | | −7.02 | −7.40 | ≤ −6.6 |
| SP 30-6 | −7.71 | | | | | |
| 04557−6753 | −6.79 | −7.47 | | −7.44 | −7.49 | |
| 05003−6712 | −7.33 | −6.73 | | −7.89 | −8.05 | |
| 05009−6616 | −7.12 | −6.45 | | −7.54 | −7.59 | |
| 05099−6740 | −6.68 (?) | −6.04(?) | | −7.93 | −8.53 | |
| 05112−6755 | — | −6.17 | −6.80 | −7.27 | −7.23 | |
| 05112−6739 | — | −6.23 | | −7.52 | −7.31 | |
| LI-LMC0578 | −7.51 | −7.43 | | −7.85 | −7.76 | |
| 05128−6455 | −7.64 | −7.26 | | −7.82 | −7.91 | |
| 05190−6748 | — | −6.79 | | −7.03 | −7.21 | |
| TRM088 | −7.06 | −7.01 | | −7.44 | −7.66 | |
| TRM045 | −7.31 | | | −7.51 | | |
| 05291−6700 | −8.33 | −7.58 | | −8.27 | −8.61 | |
| 05294−7104 | −7.31 | −6.52 | | −7.81 | −8.51 | |
| 05295−7121 | −7.16 | −6.94 | | −7.55 | −7.77 | |
| 05298−6957 | −6.58 | −6.74 | | −7.26 | −7.19 | ≤ −6.1 |
| 05300−6651 | −7.18 | −7.24 | | −7.45 | −7.63 | |
| 05316−6605 | −7.92 | −7.07 | | | −8.77 | |
| 05329−6709 | — | −5.75 | −5.82 | −6.81 | −7.30 | ≤ −6.3 |
| 05348−7024 | — | | | −6.63 | | |
| 05360−6648 | — | | | −7.23 | | |
| 05402−6956 | −6.87 | −6.58 | | −7.29 | −7.59 | ≤ −6.5 |
| 05506−7053 | — | −6.16 | | −7.03 | −7.11 | |
| 05558−7000 | −7.55 | −6.92 | | −7.90 | −8.01 | |

The best-understood method appears to be using the CO emission (Knapp & Morris 1985, Mamon et al. 1988). This molecule is abundant and the emission is easily interpreted; it is therefore widely used for Galactic sources. However, CO emission from circumstellar envelopes is not observable at the distance of the LMC with present instruments. Jura (1986, 1987) has proposed to determine mass-loss rates for carbon stars using the observed IRAS 60μm emission. The simplicity of this approach has led to its wide-spread use, even for oxygen-rich envelopes for which it was not derived originally. Even for Galactic sources, a reliable 60μm flux is not always available, and for our LMC sources the Jura method cannot be directly applied.

There are a number of secondary methods available to estimate mass-loss rates. They are discussed in the following subsections. We will derive the *dust* mass-loss rates. For scaling to gas mass-loss rates a gas-to-dust ratio has to be assumed.

### 8.1. The 9.8μm feature

Schutte & Tielens (1989) have presented detailed calculations of the strength of the 9.8μm feature as function of mass-loss rate and show that its strength and sign (emission or absorption) can be used as a determinant. They give a formula for the gas mass-loss rate as function of the 10μm optical depth, $\tau_{10}$, which can be simplified if we assume that the velocity of the dust and the gas is equal and that the outflow velocity is constant. In terms of the dust mass-loss rate we obtain:

$$\dot{M}_{\rm dust} = 5 \times 10^{-4}\, 2\pi r_0 v_\infty \tau_{10} \quad {\rm gm\,s^{-1}} \qquad (3)$$

where $r_0$ is the inner radius of the shell, in cm, and $v_\infty$ is the outflow velocity at infinity, in km s$^{-1}$. The constant arises from the absorption strength of the 10μm feature. Taking $r_0 = 2 \times 10^{14}$ cm and $v_\infty = 10\,{\rm km\,s^{-1}}$, this becomes:

$$\dot{M}_{\rm dust} = 1 \times 10^{-8} \tau_{10} \quad {\rm M_\odot\,yr^{-1}}. \qquad (4)$$

However, converting the observed value of $A_{10}$ to an optical depth is not easy, because of the competing effects of emission and self-absorption. The feature is strongest around $\tau_{10} \approx 0.2$ (Groenewegen, private communication) and goes into absorption around $\tau_{10} \approx 3.0$. Dust mass-loss rates for the sample observed with TIMMI would thus be less than $3 \times 10^{-8}\,M_\odot\,yr^{-1}$. The curves in figure 5 of Schutte & Tielens can also not be used for converting $A_{10}$ to a mass-loss rate; the slope of the relation is very shallow and only two of the observed points fall near a model curve.

### 8.2. The near–mid-infrared colour temperature

A correlation between the L–[12] colour (or related quantities) and the mass-loss rate derived with different methods, has been noted by various authors (Nguyen-Q-Rieu et al. 1979, Jones et al. 1983, Baud & Habing 1983, Schutte & Tielens 1989, Whitelock et al. 1994). The first two papers find it as a relation between the OH luminosity and the infrared colours.

Whitelock et al. (1994) calculate mass-loss rates by scaling the Jura formalism to the $25\mu m$ IRAS flux density using scaling factors which differ for oxygen and carbon stars (see also Wood et al. 1992). For a large sample of Galactic AGB stars, they find a tight correlation between these mass-loss rates and the K–[12] colour. The relation breaks down at dust mass-loss rates above $5 \times 10^{-8}\,M_\odot\,yr^{-1}$ where the modification of the Jura equation is no longer valid. Since K–[12] is well determined for our sample, we can apply this to our sources. The relation can be quantified as:

$$\log(\dot{M}_{\mathrm{dust}}) = 0.57 \times (K - [12]) - 10.35 \quad (5)$$

where an expansion velocity of $10\,km\,s^{-1}$ is used, which is more typical for Miras than the $15\,km\,s^{-1}$ used by Jura (1987); the latter value may be more appropriate for high-mass-loss AGB stars. The equation applies for $K - [12] < 5.5$.

Table 9 lists the resulting values for our sample of AGB stars. Dashes indicate that the colours fall far outside the range of applicability.

Schutte & Tielens (1989) derive a similar relation from their modeling described above, in terms of the infrared colour temperature, measured between 3.8 and $12.5\mu m$, and the mass-loss rate. Their relation extends to much higher mass-loss rates than the Whitelock et al. relation. From their figure 5 we estimate the relation to be:

$$\log(\dot{M}_{\mathrm{dust}}) = 0.58 \times (L' - [12]) - 8.97 \quad (6)$$

for dust mass-loss rates between $2 \times 10^{-8}$ and $5 \times 10^{-7}\,M_\odot\,yr^{-1}$. For higher values this approximation somewhat overestimates the result from Schutte & Tielens. Table 9 lists the derived values using the IRAS $12\mu m$ flux. However, the Schutte & Tielens result uses the true dust continuum at $12\mu m$ while the IRAS $12\mu m$ flux includes a contribution from the silicate feature. This will lead to an overestimate of the mass-loss rate, which, judging from the N–[9.8] colours of the TIMMI sample, may be as high as a factor of four. We note that the calibration of Whitelock et al. uses the broadband IRAS $12$-$\mu m$ magnitude. A separate column gives the result for the TIMMI sources using the [9.8] flux corrected for $A_{10}$. This correction significantly improves the agreement between the two methods.

### 8.3. The near-infrared optical depth

An alternative method for determining mass-loss rates is via the optical depth of the envelope derived from near-infrared colours. This method was developed by Le Bertre (1987,1988a, 1988b; Epchtein et al., 1990), who modeled several variable carbon stars solving the radiative-transfer equation. The model uses a relation between the opacity at $1\mu m$ and the colours H–K and K–L' (Epchtein et al., 1990, Fig. 5). In the model it is assumed that: (1) the star emits like a blackbody with $T_\star = 2200$ K; (2) the dust envelope is spherical; (3) the density law is $n(r) \propto r^{-2}$ and the expansion velocity is constant, corresponding to a constant mass-loss rate; (4) the inner radius of the dust shell, $R_i$, is such that the grain temperature at $R_i$ is the condensation temperature for carbon grains, $T_c = 900 - 1000$ K; (5) the dust emissivity law follows $Q(\lambda) \propto \lambda^\beta$. This is in principle only valid for carbon stars. The value of $\beta$ which fits the energy distribution of carbon stars in the region 1–$30\mu m$ is $\beta = -1.0$. (e.g. Rouleau & Martin 1991). Le Bertre used $\beta = -1.3$ as suggested by Jura (1983). Assumption (2), (3) and (5) are also used in the Jura formalism.

The numerical relation between $\tau_{1\mu m}$ and H–K and K–L' is given in Table 10. We can now calculate the dust mass-loss rate using the following equations:

$$R_\star^2 = L_\star/(4\pi\sigma T_\star^4) \quad (7)$$

$$R_i^2 \int Q_\lambda B_\lambda(T_c)\,d\lambda = \frac{(R_\star^2)}{2} \int Q_\lambda B_\lambda(T_\star)\,d\lambda \quad (8)$$

$$\dot{M}_{\mathrm{d}} = 16\pi a\rho\,\tau_{1\mu m}\,R_i V_e/\,3Q(1\mu m) \quad (9)$$

with the following assumptions: $R_o \gg R_i$ (used in the third formula; $R_o$ is the outer radius of the dust shell); $T_c = 950$ K; $Q(\lambda_{1\mu m}) = 0.19$; $a = 0.1\,\mu m$ (grain size); $\rho = 3\,g\,cm^{-3}$, which is valid for carbon-rich dust (for silicates $\rho = 2.7$); $V_e = 10\,km\,s^{-1}$. The derived values can be directly scaled to accommodate different values of $Q(1\mu m)$, the grain size, $\rho$, or the distance, without the need for numerical calculations.

**Table 10.** Optical depth and near-infrared colours

| $\tau_{1\mu m}$ | H–K | K–L' |
| --- | --- | --- |
| .0 | .84 | 0.91 |
| .1 | .88 | 1.01 |
| .14 | .89 | 1.05 |
| .2 | .91 | 1.10 |
| .28 | .94 | 1.16 |
| .4 | .98 | 1.25 |
| .57 | 1.03 | 1.36 |
| .8 | 1.10 | 1.49 |
| 1.13 | 1.20 | 1.66 |
| 1.6 | 1.33 | 1.86 |
| 2.26 | 1.51 | 2.10 |
| 3.2 | 1.75 | 2.40 |
| 4.53 | 2.05 | 2.84 |
| 6.4 | 2.46 | 3.32 |
| 9.05 | 3.05 | 3.96 |
| 12.8 | 3.85 | 4.80 |

Figure 10 shows the fit of the Le Bertre track to our sample. It gives a fair fit, with a tendency for the sources to have a slightly too low K−L. This may be caused either by molecular opacities distorting the spectrum, or the spread may indicate a range in stellar temperatures. We conclude that the colours are consistent with reddened black bodies and that the Le Bertre formalism can in principle be applied.

We can approximate the above formulae as:

$$\dot{M}_{\rm d} = 7.53 \times 10^{-11}\, \tau\, L_\star^{0.5} \qquad (10)$$

in solar units. The resulting values are listed in Table 9. They are generally about 2.5 times lower than derived with the previous method. The same effect is seen if we use the sample of Galactic-cap Miras of Whitelock et al. (1994), where the HKL colours would indicate $\tau = 0$. A possible explanation is that the central-star temperatures are higher than assumed here, of the order of 3000 K rather than 2200 K. There are several other uncertainties inherent in the above simplifications. The most important may be that Eq. (10) is derived for carbon stars, whereas the majority of the stars in our sample may be oxygen rich. However, the Le Bertre track could give a good approximation to the *relative* mass-loss rates of the stars in our sample. We could in fact re-calibrate it using the Whitelock relation, which would imply adding roughly 0.4 to $\log \dot{M}$ derived from the Le Bertre method.

We note that this relation implies that the observational H−K versus K−[12] relation discussed in the previous section in fact traces the mass loss.

### 8.4. OH luminosity

Baud & Habing (1983) have shown that there is a correlation between OH-maser strength and mass-loss rate. This is in fact to be expected if the masers are saturated and operate at maximum pumping efficiency. The maser is pumped via a transition at a wavelength of $35\mu$m, and the theoretical efficiency for converting $35\mu$m photons to OH maser photons is 25%. In this case the maser 'measures' the infrared continuum at $35\mu$m and the Jura formula applies with a wavelength shift from $60\mu$m to $35\mu$m. It is noteworthy that the maser strength is determined by the infrared flux and not by the number of OH molecules. Thus, the OH luminosity is dependent on the *dust* mass-loss rate and not the OH-gas mass-loss rate! This method was used by Wood et al. (1992) to derive mass-loss rates for their OH-emitting stars in the LMC.

The calibration by Baud & Habing was done with a small sample and has not been refined. Many more Galactic Miras with OH emission are now known, and a re-investigation of the method is possible. This is especially important because the newly discovered OH-emitting stars in general show pump efficiencies below the theoretical value of 25%.

A catalogue of OH-emitting stars can be found in te Lintel Hekkert et al. (1991). We have cross-correlated this catalogue with the Galactic CO catalogue of Loup et al. (1993) which also gives mass-loss rates derived from the CO emission. The distance is calculated by assuming a bolometric luminosity of $10^4\,\mathrm{L}_\odot$.

Fig. 11a–c shows a comparison of the different methods to calculate mass-loss rates for these Galactic stars. The filled circles indicate sources for which $L_{\rm OH} > 200\,\mathrm{Jy\,kpc}^2$ which is the OH luminosity of the faintest detected sources in the LMC.

(The five stars are 26.5+0.6, 30.1−0.7, 32.8−0.3, 104.9+2.4, 127.9−0.0; 32.8−0.3 has no $60\mu$m detection.) The correlation with the CO mass-loss rates is poor. For the brightest OH emitters, the CO implies much too low mass-loss rates. This has been noted before (Heske et al. 1990). Van der Veen & Rugers (1989) suggest that this occurs when the CO excitation line at $4.6\mu$m becomes optically thick, but this may not be the correct explanation since collisional excitation is dominant over radiative excitation. Alternatively, Schutte & Tielens (1989) discuss the possibility that the mass loss was lower in the past, and that the CO emission still is dominated by material far from the star dating back to earlier times. Recently, Groenewegen (1994) has found evidence for this in the case of OH32.8−0.4. In either case the Jura method would give better values for the present-day mass loss than would the CO emission.

The OH mass-loss rates, calculated as in Baud & Habing, show a general correlation with the Jura values, however there is significant disagreement at the higher values. The scatter is in part due to the range in pump efficiency: calculating the ratio of the $25\mu$m IRAS flux of this sample and the OH peak flux densities shows efficiencies in the range of 1–20%, with large scatter. However, the disagreement at the high mass-loss rates suggests a closer look at the method.

Baud & Habing assume that the surface brightness of the OH maser is approximately constant and therefore that $L_{\rm OH} \sim (\dot{M}_d/V_e)^2$. This assumption will be incorrect if the OH strength is dependent on the dust radiation, in which case $L_{\rm OH} \sim \dot{M}_d/V_e$ (calculating $L_{\rm OH}$ as described in Baud & Habing). Figure 11d shows the observed data. Drawing a line through the higher points only, to eliminate the sources with low OH pump efficiencies, we obtain approximately:

$$\log \dot{M}_{\rm dust} = \log(L_{\rm OH} V_{\rm exp}) - 9.8. \qquad (11)$$

The uncertainties are large, because few OH emitters operate at peak efficiency. Equation (11) will therefore for most stars give lower limits to the mass-loss rates. In addition, the measurement uncertainties in the OH-peak flux densities are large for the Magellanic Cloud objects. The derived values are listed in Table 9. There is reasonable agreement with the K−[12] and L−[12] methods.

### 8.5. The $12\mu$m/$25\mu$m flux ratio

Van der Veen & Olofsson (1989) also discuss mass-loss rates derived from the ratio of the IRAS $12\mu$m and $25\,\mu$m flux. This is an attractive method which only depends on the dust temperature which in turn is determined by the optical depth of the shell. and chemical composition. However, the calculated mass-loss rates correlate poorly with values determined by other methods, most noticeably CO (Whitelock et al. 1994, Epchtein et al. 1990). There is a general trend of redder IRAS colours with higher mass-loss rates, but with large scatter. It is possible that the derived values are sensitive to long-term fluctuations in the mass-loss rate; Whitelock et al. on the other hand conclude that the optical depth through the shell is not the dominant factor controlling the 25-$\mu$m to 12-$\mu$m flux ratio.

For our sample this is aggravated by the fact the $25\mu$m flux is the least reliable of the available data points. We have therefore chosen not to apply this method. Better mid-infrared data (e.g. from ISO) is needed to investigate the applicability of the method.

*8.6. Results*

The values for the mass-loss rates determined from the ratio of near-infrared to mid-infrared emission are preferred, since these methods are quite well calibrated. The L−[12] should still be corrected for silicate emission and presently are over-estimated. The near-infrared colour method probably gives reasonable *relative* mass-loss rates but the absolute calibration is uncertain; the present results indicate that this method underestimates the mass-loss rate by a factor of 2.5 on average. For one object (05099−6740) there is doubt whether the IRAS source is associated with the star and the mass-loss rates are marked accordingly.

We have not discussed the problem of source variability. For the methods using colours, in principle simultaneous measurements are required: this is a potential problem wherever IRAS data is used. The results of Whitelock et al. (1994), who find good agreement for the K−[12] mass-loss rates with values derived from CO, suggest that in many cases the effect of variability on the IRAS fluxes is small. However, the effect may well be larger for the present sample, which contains more luminous stars. The IRAS data is likely biased towards the peak of the pulsation cycle whereas the NIR data is taken at a random phase: the near-infrared–mid-infrared methods may therefore overestimate the mass-loss rates by a factor of 2–5. More accurate mass-loss rate will require monitoring over a full period at a range of wavelengths.

There appears to be no simple way to directly determine gas mass-loss rates for Magellanic Cloud stars. We can only derive dust mass-loss rates which have to be scaled by an uncertain factor. We find a large range of dust mass-loss rates. If we assume a gas-to-dust ratio of 400, the resulting gas mass-loss rates would between $5 \times 10^{-4}$ and $4 \times 10^{-6}$ $M_\odot \text{yr}^{-1}$. The highest value is found for 05329−6709 which is also one of the brightest IRAS sources in the sample and has a very high bolometric luminosity. However, for the full sample there is no evidence of a correlation between mass-loss rates and luminosity. We also find no evidence that the mass-loss rates in the SMC are systematically lower than in the LMC. This result would not be unexpected if the highest values of $\dot{M}$ are limited by the momentum in the radiation field, as suggested by Vassiliadis & Wood (1993).

## 9. Expansion velocities

Expansion velocities are available for only four of the IRAS-detected AGB stars in the LMC, all based on the OH detections by Wood et al. (1992). These four objects can be compared to Galactic AGB stars. Wood et al. find that the OH-expansion velocities of LMC AGB stars are 40% lower than those of Galactic OH/IR stars of similar IRAS colours. This difference is particularly clear when expansion velocity is plotted against period. For this comparison, Wood et al. only use Galactic stars within 1° of the Galactic plane, selecting highly luminous sources.

We can apply the same procedure to the sample of Galactic OH/IR stars used in subsection 8.4. The comparison is shown in Fig. 12. Note that we plot expansion velocity rather than the $\Delta V$ used by Wood et al. Our Galactic sample contains five objects with OH luminosities similar to those of the MC stars, although this depends on distances which are uncertain. The upper panel shows that, compared to the Galactic star velocities, the LMC expansion velocities may indeed be lower, but not by as large a factor as found by Wood et al. The difference in conclusion is mostly caused by the fact that we use half the total width of the profile rather than half the separation of the two highest maser peaks to define the expansion velocity. The difference can be several km s$^{-1}$ in the case of noisy profiles.

The middle panel shows the comparison as a function of the IRAS colour. The five luminous Galactic stars are indicated by crosses. They show IRAS colours which are comparable to the LMC stars. The lower panel gives the expansion velocity versus period, the latter from van Langevelde et al. (1990). The LMC sources appear very similar to the luminous Galactic OH/IR stars. Their expansion velocities appear to be lower by about 10–20% although there are insufficient data to draw a firm conclusion. Higher sensitivity OH observations would be desirable to better define the expansion velocities of the LMC stars.

Almost all other MC sources in our sample have $F_{12}/F_{25}$ close to unity, which is significantly bluer than the OH-detected sources. Compared with the Galactic sources in the middle panel, we expect that their OH emission will be considerably fainter than the detection limit of the Parkes observations (Wood et al. 1992).

## 10. Period determination

Two of the sources in our sample have been monitored at SAAO in the past: HV 12956 (Whitelock et al. 1989) and 04407−7000 (unpublished). 04407−7000 was discovered and monitored as part of a program on AGB stars in the Galactic cap; it was later dropped because of its probable association with the LMC. 05329−6709 was monitored by Wood et al. (1992), who derive a period of 1260 days. In addition, Wood et al. give periods for a number of other IRAS stars in the Magellanic Clouds.

HV 12956 is a known optical variable, for which a period of 517 days has been derived (Payne-Gaposchkin & Gaposchkin 1966). Its spectral type is M5e at maximum which is late for an SMC object (Whitelock et al. 1989). The present observations confirm the association of the IRAS source with the Harvard variable: Whitelock et al. considered this as less certain because the colours of the star are not obviously reddened.

Fig. 13 shows the variation of the infrared magnitudes including both the previous and the new observations. The most recent SAAO observations of 1993 November and the IRAC photometry are also included. For HV 12956 there is sufficient data for an unambiguous determination of the period. Using a program written by Luis Balona, we obtain 515 days, with a nominal uncertainty of 2 days. A period can also be derived by comparing the JD at maximum given by Payne-Gaposchkin & Gaposchkin, JD 2431436, with the best-determined infrared maximum at JD 2447100. This gives a period of $522 \pm 1$ days which should be regarded as an upper limit because the infrared maximum typically lags 0.1 to 0.2 of a cycle behind the optical maximum (Lockwood & Wing 1971; Strecker 1973). For 04407−7000 there are two possible periods: 1220 days or 770 days. Further monitoring is needed before either of these can be excluded. Both sources have long periods compared to optical Miras (e.g. Whitelock et al. 1991) and are similar to the Wood et al. sources. They also have bright bolometric magnitudes ($M_{\rm bol} = -7.2$ for 04407−7000 and $M_{\rm bol} = -6.7$ for HV 12956), consistent with the suggestion that long periods are only reached for stars with high-mass progenitors.

For HV 12956 a longer modulation is also clear from Fig. 14. Neither the time scale for this modulation, nor its regularity, can be determined from the available data. Such modulations are commonly seen in long-period Miras. It is possible that similar, but smaller amplitude, changes occur in shorter-period Mira where they are less obvious.

Optical Miras show a strong period–luminosity relation, with higher luminosity coinciding with longer periods. The relation is quite narrow in the K-band; it shows more scatter when the bolometric luminosity is used. A discussion of the period–luminosity diagram for LMC Miras is given by Feast et al. (1989), based on stars with periods less than 420 days. Figure 14 shows their data together with the present sample. Clearly the long-period sources are more luminous than would be predicted by extrapolating from the optical variables. This is consistent with other long-period variables in the LMC (Feast et al. 1989; Wood, Bessel & Fox 1983; Wood et al. 1992; see Fig. 20 in Vassiliadis & Wood 1993). The $PL$ relation used by Wood (1990) and Vassiliadis & Wood is aimed at fitting these longer-period stars and therefore is significantly steeper than commonly-used $PL$ relations: it does not fit the shorter-period variables well (Groenewegen 1994). Feast et al. (1989) show that a PLC relation using the J−K colour fits the short-period variables better. This can not be directly used for the long-period variable stars where the J−K colour is affected by circumstellar extinction.

The fit to the carbon-rich stars in Figure 14 is clearly different from that to the oxygen-rich stars. Glass et al. (1987) have suggested that the high molecular opacities of carbon atmospheres in the J and H band causes the bolometric luminosity to be underestimated. In addition, inter-band water vapour absorption in oxygen Miras may cause their luminosity to be slightly overestimated. This effect should be understood before carbon and oxygen Miras can be compared. We note that the 'peculiar' source HV 12956 falls about two magnitudes above the relation. If it is indeed a foreground object, it would still have to be at a distance of around 20 kpc and 10 kpc above the Galactic plane. An association with the SMC itself seems more likely.

The PL relation has been interpreted as an evolutionary sequence, but this is unlikely for several reasons: (1) the Mira life times ($\sim 10^5$ yr) are too short to give an appreciable evolution in luminosity; (2) Miras in individual globular clusters show a very small range in luminosities; (3) For Galactic Miras the velocity dispersion is a function of period (Feast 1963). An alternative suggestion is that the long-period variables with periods larger than 1000 days originate from a change from overtone to fundamental pulsation mode. However, this would result in a horizontal translation in Figure 14 while we observe that the stars with the longest periods are shifted upward with respect to shorter-period stars. Thus, the likely explanation for the long-period MC stars is that they originate from higher-mass progenitor stars (e.g. Feast 1989) (see also the discussion in Whitelock et al. 1991). The fact that long periods as shown by the MCs Miras are not found in the outer bulge (Whitelock et al. 1991), where a population with massive progenitors is not present, also favours that long periods arise from a young, relatively high-mass stellar population.

The evolutionary models of Vassiliadis & Wood (1993) explain the observed distribution of periods with luminosity quite well: here the spread at any given luminosity is largely caused by the fact the mass of the star decreases due to mass loss, increasing its period. This effect is largest for massive stars and could be the cause of the absence of a clearly-defined $PL$ relation for AGB stars with long periods. The evolutionary tracks in of Vassiliadis & Wood (1993) in the $PL$ diagram are inclined to the observed $PL$ relation, following a much shallower angle. This is in very good agreement with a $PL$ relation found for globular clusters, connecting semi-regular variables with the Miras (Whitelock 1986) with as only difference that the observed globular-cluster relation is shifted to shorter periods with respect to the theoretical predictions (Zijlstra 1995). It seems likely that this shallow relation is the true evolutionary sequence, and that the observed $PL$ relation of Fig. 14 mainly traces a range of initial masses.

## 11. Conclusions

We have presented the results of an infrared survey of IRAS sources in the Magellanic Clouds. We find a significant number of obscured AGB stars, constituting an additional AGB population to the low-mass-loss, short-period Miras already known. We present positions and near-infrared photometry. The stars are highly luminous, with luminosities in general between $\log L = 4.0$ and $\log L = 4.7$. 10-$\mu$m photometry in several bands has been done for a number of these sources; the results indicate that oxygen-rich shells dominate. Two of the sources show a very strong silicate feature, which in our Galaxy is only found for low-luminosity IRAS sources. It is shown that oxygen and carbon stars can be separated on the basis of near-infrared–mid-infrared colours: the relation is valid both for Galactic and Magellanic Cloud stars. However, the uncertainties in the IRAS data are too large to allow one to directly use this relation in the MCs; instead, ground-based 10-$\mu$m data should be used.

We discuss several methods to obtain mass-loss rates. The gas mass-loss rate cannot be directly determined for these sources and we have only been able to derive dust mass-loss rates. The K−[12] and L−[12] colours give the preferred estimates of the mass-loss rates, since these methods are well calibrated against CO mass-loss rates and correlate well with those. In principle the near-infrared colours allow one to measure the optical depth through the shell, however the models yield only relative mass-loss rates. We give a new relation to derive approximate mass-loss rates from the OH luminosity.

The LMC sources appear to show slightly lower expansion velocities than Galactic OH/IR stars, although the difference may not be as large as found by Wood et al. (1992). Overall the obscured AGB stars in the Magellanic Clouds are quite similar to the (relatively few) luminous Galactic OH/IR stars.

For two of the sources we are able to derive periods. For several other objects, periods were already known. All periods are very long ($P \gtrsim 1000$ days), with one exception (the SMC source HV 12956), and all stars fall above the period–luminosity relation as derived for short-period Miras. It is like-ly that the short-periods and long-periods AGB stars have different progenitor stars and are not evolutionary related, where long periods are derived from high-mass progenitors. The sensitivity limit of IRAS is insufficient to detect the circumstellar envelopes of low-luminosity AGB stars. ISO observations will be needed to study the mass loss of these stars. An analysis of how the mass loss depends on the characteristics of the progenitor stars will require such ISO observations.

**Acknowledgements**

We have made use of the infrared archive at the SAAO for this research and of the SIMBAD database. We thank Robin Catchpole and Michael Feast for the earlier observations of HV 12956 and 04407−7000 used in this paper. We would also like to thank the supporting staff at ESO and SAAO, in particular Andrea Monetti, Hans-Ulrich Käufl and Francois van Wijk, for their help during the observations. We thank Xander Tielens for sending us the data used in Schutte & Tielens (1989), and Bruno Leiblundgut for the critical reading of the manuscript.

**References**

Baud B., Habing H.J., 1983, A&A, 127, 73
Bessell M.S., Brett J.M., Scholz M., Wood P.R., 1989a, A&A, 213, 209
Bessell M.S., Brett J.M., Scholz M., Wood P.R., 1989b, A&ASS, 77, 1
Blanco V.M., McCarthy M.F., Blanco B.M., 1980, ApJ, 242, 938.
Blöcker T., Schönberner D., 1991, A&A, 244, L43
Bouchet P, Moneti A., Slezak E., Le Bertre T., Manfroid J., 1989, A&ASS, 80, 379
Carter B.S., 1990, MNRAS, 242, 1
Chiosi C., Maeder A., 1986, ARAA, 24, 329
Elias J.H., Frogel J.A., Schwering P.B.W., 1986, ApJ, 302, 675
Epchtein N., Le Bertre T., Lépine J.R.D., 1990, A&A, 227, 82
Feast M.W., 1963, MNRAS, 125, 367
Feast M.W., Robertson B.S.C., Catchpole R.M., Lloyd Evans T., Glass I.S., Carter B.S., 1982, MNRAS, 201, 439
Feast M.W, Walker A.R., 1987, ARA&A, 25, 345
Feast M.W., Glass I.S., Whitelock P.A., Catchpole R.M., 1989, MNRAS, 241, 375
Feast M.W., Whitelock P.A., Carter B.S., 1990, MNRAS, 247, 227
Fouqué P., Le Bertre T., Epchtein N., Guglielmo F., Kerschbaum F., 1992, A&ASS, 93, 151
Glass I.S, Catchpole R.M., Feast M.W., Whitelock P.A., Reid I.N., 1987, in: Late Stages of Stellar Evolution. Eds. Kwok S., Pottasch S.R. (Reidel, Dordrecht) p.51
Groenewegen M.A.T., 1994, A&A, 290, 544
Groenewegen M.A.T., 1995, in: proc. conf. Circumstellar Matter, Edinburgh, Astrophysics and Space Science, eds. G.D. Watt and P.M. Williams (Kluwer, Dordrecht), p.321
Guglielmo F., Epchtein N., Le Bertre T., Fouqué P., Hron J., Kerschbaum F., Lépine J.R.D, 1993, A&ASS, 99, 31
Han Z., Podsialowski Ph., Eggleton P.P., 1994, MNRAS, 270, 121
Hearn A.G., 1989, in "From Miras to Planetary Nebulae: Which Path for Stellar Evolution?", eds. M.O. Mennessier and A. Omont (Editions Frontières, Gif sûr Yvette), p121
Heske A., Forveille T., Omont A., van de Veen W.E.C.J., Habing H.J., 1990, A&A, 239, 173
Hughes S.M.G., Wood P.R., 1990, AJ, 99, 784
Jones T.J., Hyland A.R., Gatley I., 1983, ApJ, 273, 660
Jura M., 1983, ApJ, 267, 647
Jura M., 1986, ApJ, 303, 327
Jura M., 1987, ApJ, 313, 743
Käufl U., Jouan R., Lagage P.O., Masse P., Mestreau P. Tarrius A., 1992, The Messenger, 70, 67
Knapp G.R., Morris M., 1985, ApJ, 292, 640
Le Bertre T., 1987, A&A, 176, 107
Le Bertre T., 1988a, A&A, 190, 79
Le Bertre T., 1988b, A&A, 203, 85
Le Bertre T., 1993, A&ASS, 97, 729
Lockwood G.W., Wing R.F., 1971, ApJ, 169, 63
Lloyd Evans T., 1984, MNRAS, 208, 447
Loup C., Zijlstra A.A., Waters L.B.F.M., 1995, A&ASS, accepted for publication (paper I)
Loup C., Forveille T., Omont A., Paul J.F., 1993, A&ASS, 99, 291
Mamon G.A., Glassgold A.E., Huggins P.J., 1988, ApJ, 328, 797
McGregor P.J., 1994, PASP, 106, 508
Nguyen-Q-Rieu, Laury-Micoulaut C., Winnberg A., Schultz G.V., 1979, A&A, 75, 351
Payne-Gaposchkin C., Gaposchkin S., 1966, Smithson. Contr. Astrophys., 9, 1
Pottasch S.R., Bignell C., Olling R., Zijlstra A.A., 1988, A&A, 205, 248
Rebeirot E., Martin N., Mianes P., Prévot L., Robin A., Rousseau J., Peyrin Y., 1983, A&ASS, 51, 277
Reid I.N., 1991, ApJ, 382, 143
Reid I.N., Tinney C.G., Mould J.R., 1990, ApJ, 348, 98
Reid I.N., Glass I.S., Catchpole R.M., 1988, MNRAS, 232, 53
Robertson B.S.C., Feast M.W., 1981, MNRAS, 196, 111
Rouleau F., Martin P.G., 1991, ApJ, 377, 526
Sandulaek N., 1970, Contr. Cerro-Tololo Obs., 89
Sandulaek N., Philip A.G., 1977, Pub. Warner and Swasey Obs., 2, No. 5
Schutte W.A., Tielens A.G.G.M., 1989, ApJ, 343, 369
Schwering P.B.W., 1989, A&ASS, 79, 105
Schwering P.B.W., Israel F.P., 1989, A&ASS, 79, 79
Schwering P.B.W., Israel F.P., 1989, A catalog of IRAS sources in the Magellanic Clouds (Kluwer, Dordrecht)
Shore S.N., Sandulaek N., 1984, ApJSS, 55, 1
Smith V.V., Lambert D.L., 1989, ApJ, 345, L75
Smith V.V., Lambert D.L., 1990, ApJ, 361, L69
Smith V.V., Plez B., Lambert D.L., Lubowich D.A., 1994, ApJ, 441, 735
Strecker D.W., 1973, University of Minnesota Astrophysics Report, no. 15
te Lintel Hekkert P., Caswell J.L, Habing H.J., Haynes R.F., Norris R.P., 1991, A&ASS, 90, 327
Tinney C.G., Mould J.R., Reid I.N., 1993, AJ, 105, 1045
van der Veen W.E.C.J., Olofsson H., 1989, in "From Miras to Planetary Nebulae: Which Path for Stellar Evolution?", eds. M.O. Mennessier and A. Omont (Editions Frontières, Gif sûr Yvette), p139
van der Veen W.E.C.J., Rugers M., 1989, A&A, 226, 183
van Langevelde H.J., van der Heiden R. van Schooneveld C., 1990, A&A, 239, 193
Vassiliadis E., Wood P.R., 1993, ApJ, 413, 641
Weideman V., 1987, A&A, 188, 74
Westerlund B.E., Olander N., Hedin B., 1981, A&ASS, 43, 267
Whitelock P.A., 1986, MNRAS, 219, 525
Whitelock P.A., Feast M.W., Menzies J.W., Catchpole R.M., 1989, MNRAS, 238, 769
Whitelock P.A., Feast M.W., Catchpole R.M., 1991, MNRAS, 248, 276
Whitelock P.A., Menzies J.W, Feast M.W., Marang F., Carter B.S.,

Roberts G., Catchpole R.M., Chapman J.W., 1994, MNRAS, 267, 711
Wood P.R., Bessell M.S., Fox M.W., 1983, ApJ, 272, 99
Wood P.R., Bessell M.S., 1985, PASP, 97, 681
Wood P.R., Whiteoak J.B., Hughes M.G., Bessell M.S., Gardner F.F., Hyland A.R., 1992, ApJ, 397, 552

Zijlstra, A.A., 1995, in: proc. conf. Circumstellar Matter, Edinburgh, Astrophysics and Space Science, eds. G.D. Watt and P.M. Williams (Kluwer, Dordrecht), p.309
Zijlstra A.A., van Hoof P.A.M., Chapman J.M., Loup C., 1994, A&A, 290, 228

**Figure captions**

**Figure 1** Contour plots of the area around 05329−6709 at J and K.

**Figure 2** J−H versus H−K for the sources observed at ESO and SAAO. The lower diagram shows an enlargement of the bottom-left corner, indicating where various classes of stars would fall. In the lower diagram, the open circles indicate possible reddened AGB stars, the filled triangles possible M-giants or supergiants, and the open stars foreground stars.

**Figure 3** Colour–magnitude diagrams for the stars in the Magellanic Clouds. IRAS-detected AGB stars from other papers are also included.

**Figure 4** J−H versus H−K for the IRAS-detected AGB stars from other papers. The filled circles indicate the stars for which Wood et al. (1992) detected OH emission.

**Figure 5** Spectral energy distribution for 20 supergiants and three objects with uncertain classification in the Magellanic Clouds. The curve in one spectrum shows a blackbody with a temperature of 2500 K.

**Figure 6** Spectral energy distribution for candidate AGB stars in the Magellanic Clouds.

**Figure 7** Mid-infrared colours for Galactic AGB stars taken from the IRAS LRS database. The cross indicates our calibrator $\gamma$ Ret. The triangles with error bars represent the TIMMI data for Magellanic Cloud stars.

**Figure 8a** Relation between $A_{10}$ and the N−[9.8] colour for Galactic AGB stars, derived from spectra taken from the IRAS LRS database.

**Figure 8b** Relation between IR colour temperature $T_c$ and $A_{10}$. Triangles indicate Galactic stars from Schutte & Tielens 1989, and the star symbols indicate LMC stars. The tracks is a representative model from Schutte & Tielens.

**Figure 9** K−[12] versus H−K. The upper diagram shows Galactic AGB stars and indicates two separate branches for oxygen-rich and carbon-rich stars. The lower diagram shows the data for the MC stars. The filled circles indicate the sources observed with TIMMI.

**Figure 10** K−L versus H−K for the MCs stars. For comparison the Galactic Cap Miras from Whitelock et al. (1994) are also shown. The line indicates the reddening track calculated by Le Bertre.

**Figure 11** Comparison between dust mass-loss rates derived from CO, 60$\mu$m flux and OH luminosity, for a sample of Galactic OH/IR stars. The filled circles indicate Galactic stars with $L_{\rm OH} > 200\,{\rm Jy\,kpc}^{-2}$ which would have been detectable at the distance of the LMC. The 60$\mu$m emission is assumed to give the most reliable value. The OH mass-loss rate is calculated as in Baud and Habing (1983). Panel d shows a relation between $L_{\rm OH}$ and the 60$\mu$m mass-loss rate. The line gives our proposed calibration for OH masers which operate at maximum pumping efficiency.

**Figure 12** Expansion velocities of LMC and Galactic OH/IR stars. The crosses indicate Galactic OH/IR of comparable OH luminosity to the LMC sources.

**Figure 13** Monitoring data for HV12956 and 04407−7000. The latter require more data to allow an unambiguous period determination.

**Figure 14** PL relation for Magellanic Cloud Miras. The horizontal line indicates two possible periods for a single object. The dashed lines show the relations proposed by Feast et al. (1989) for oxygen and carbon Miras; the dotted line shows the calibration for Galactic sources of Whitelock et al. (1991).